\documentclass[12pt,my_subfigure,onecolumn,a4]{IEEEtran}
\usepackage{multirow}

\input{epic.sty}
\input{eepic.sty}

\newcommand{\subpostscript}[2]{
    \setlength{\epsfxsize}{#2}
    \epsfbox{#1}
}

\input{psfig}               % for including ps file
\input{epsf}                % for including eps file

\newcommand{\bthm} {\begin{thm} }
\newcommand{\ethm} {\end{thm}}
\newcommand{\blem} {\begin{lem} }
\newcommand{\elem} {\end{lem}}
\newcommand{\bcor} {\begin{cor} }
\newcommand{\ecor} {\end{cor}}

\newcommand{\beq}  {\begin{equation}}
\newcommand{\eeq}  {\end{equation}}
\newcommand{\beqa}{\begin{eqnarray}}
\newcommand{\eeqa} {\end{eqnarray}}
\newcommand{\bdis} {\begin{displaymath}}
\newcommand{\edis} {\end{displaymath}}
\newcommand{\bitem}{\begin{itemize}}
\newcommand{\eitem}{\end{itemize}}
\newcommand{\ba}  {\begin{array}}
\newcommand{\ea}  {\end{array}}

\newcommand{\mod} {\mathrm{mod}}

\newcommand{\bmA} {\mbox{\boldmath $A$}}
\newcommand{\bmB} {\mbox{\boldmath $B$}}
\newcommand{\bmC} {\mbox{\boldmath $C$}}

\newcommand{\bmD} {\mbox{\boldmath $D$}}
\newcommand{\bmE} {\mbox{\boldmath $E$}}
\newcommand{\bmG} {\mbox{\boldmath $G$}}
\newcommand{\bmR} {\mbox{\boldmath $R$}}

\newcommand{\bmI} {\mbox{\boldmath $I$}}
\newcommand{\bmX} {\mbox{\boldmath $X$}}

\newcommand{\bmw} {\mbox{\boldmath $w$}}

\newcommand{\bmn} {\mbox{\boldmath $n$}}

\newcommand{\bms} {\mbox{\boldmath $s$}}
\newcommand{\bmsspt} {{\scriptsize\mbox{\boldmath $s$}}}
\newcommand{\bmx} {\mbox{\boldmath $x$}}
\newcommand{\bmy} {\mbox{\boldmath $y$}}

\newcommand{\bmzero} {\mbox{\boldmath $0$}}

\newtheorem{theorem}{Theorem}
\newtheorem{thm}[theorem]{Theorem}
\newtheorem{lemma}{Lemma}
\newtheorem{lem}[lemma]{Lemma}

\renewcommand{\QED}{\QEDopen}

\begin{document}

\begin{titlepage}

\title{{\LARGE{High Data-Rate Single-Symbol ML
Decodable Distributed STBCs for Cooperative Networks}}}

\author{\vspace{1cm}Zhihang~Yi and Il-Min~Kim\\
\vspace{5mm}
Department of Electrical and Computer Engineering\\
Queen's University\\
Kingston, Ontario, K7L 3N6\\
Canada\\
\vspace{5mm} Email: ilmin.kim@queensu.ca}

\maketitle

\begin{center}
Submit to {\it IEEE Trans. Inform. Theory as a Correspondence}
\end{center}
\vspace{5mm}

\begin{abstract}
High data-rate Distributed Orthogonal Space-Time Block Codes
(DOSTBCs) which achieve the single-symbol decodability and full
diversity order are proposed in this paper. An upper bound of the
data-rate of the DOSTBC is derived and it is approximately twice
larger than that of the conventional repetition-based cooperative
strategy. In order to facilitate the systematic constructions of
the DOSTBCs achieving the upper bound of the data-rate, some
special DOSTBCs, which have diagonal noise covariance matrices at
the destination terminal, are investigated. These codes are
referred to as the row-monomial DOSTBCs. An upper bound of the
data-rate of the row-monomial DOSTBC is derived and it is equal to
or slightly smaller than that of the DOSTBC. Lastly, the
systematic construction methods of the row-monomial DOSTBCs
achieving the upper bound of the data-rate are presented.
% for any numbers of any values of $N$ and $K$, where $N$ is
%the length of the signal vector transmitted from the source
%terminal and $K$ is the number of relay terminals.

\end{abstract}


\end{titlepage}

\section{Introduction}\label{sec:intro}

It is well-known that relay terminal cooperation can improve the
performance of a wireless network considerably
\cite{sendonaris1}--\cite{laneman2}. The basic idea of cooperative
networks is that several single-antenna terminals form a
distributed multi-antenna system by cooperation. Specifically, a
source terminal, several relay terminals, and a destination
terminal constitute a cooperative network, where the relay
terminals relay the signals from the source terminal to the
destination terminal. Because the destination terminal may receive
different signals from several relay terminals simultaneously,
some mechanism is needed to prevent or cancel the interference
among these signals.

A simple solution is the so-called {\it repetition-based
cooperative strategy}, which was proposed in \cite{laneman1}. In
this strategy, only one relay terminal is allowed to transmit the
signals at every time slot. Consequently, no interference exists
at the destination terminal, and hence, the decoding process is
single-symbol Maximum Likelihood (ML) decodable.\footnote{A code
or a scheme is said to be single-symbol ML decodable, if its ML
decoding metric can be written as a sum of several terms, each of
which depends on at most one transmitted symbol \cite{khan}.}
Furthermore, it has been shown that the repetition-based
cooperative strategy could achieve the full diversity order $K$,
where $K$ is the number of relay terminals. Due to its
single-symbol ML decodability and full diversity order, the
repetition-based cooperative strategy was used and studied in many
literatures \cite{laneman2}, \cite{anghel1}--\cite{chen2}.
However, because only one relay terminal is allowed to transmitted
the signals at every time slot, the repetition-based cooperative
strategy can be seen as a repetition code, and hence, it has very
poor bandwidth efficiency. It is easy to see that the
data-rate\footnote{In this paper, the data-rate of a cooperative
strategy or a distributed space-time code is defined as the
average number of symbols transmitted by the relay terminals per
time slot, i.e. its value is equal to the ratio of the number of
transmitted symbols to the number of time slots used by the relay
terminals to transmit all these symbols.} of the repetition-based
cooperative strategy is $1/K$.

Recently, many researchers noticed that the use of {\it
distributed space-time codes} could improve the bandwidth
efficiency of cooperative networks. In
\cite{laneman3}--\cite{yang}, the authors proved that the
distributed space-time codes had higher bandwidth efficiency than
the repetition-based cooperative strategy from the information
theory aspect. Later on, many practical distributed space-time
codes were proposed \cite{gamal}--\cite{jing2}. However, none of
those codes were single-symbol ML decodable. In \cite{hua}, Hua
{\it et al.} investigated the use of the generalized orthogonal
designs in cooperative networks.
%It is well-known that OSTBCs
%have much better bandwidth efficiency than the repetition code
%\cite{liang1,liang2,su,wang}.
It is well-known that the generalized orthogonal designs can
achieve single-symbol decodability and full diversity
\cite{alamouti,tarokh}. However, when the generalized orthogonal
designs were directly used in cooperative networks, the
orthogonality of the codes was lost, and hence, the codes were not
single-symbol ML decodable any more \cite{hua}. Very recently,
Jing {\it et al.} used the existing orthogonal and
quasi-orthogonal designs in cooperative networks and showed that
they could achieve the full diversity order \cite{jing3}. But, the
codes proposed in \cite{jing3} were not single-symbol ML decodable
in general. To the best of our knowledge, high data-rate
distributed space-time codes which achieve both the single-symbol
ML decodability and the full diversity order have never been
designed. This motivated our work.

In this paper, we propose a new type of distributed space-time
codes, namely \emph{Distributed Orthogonal Space-Time Block Codes}
(DOSTBCs), for the amplify-and-forward cooperative networks. The
proposed DOSTBCs achieve the single-symbol ML decodability and
full diversity order. An upper bound of the data-rate of the
DOSTBC is derived. Compared with the data-rate of the
repetition-based cooperative strategy, the data-rate of the DOSTBC
is approximately twice higher. However, systematic construction of
the DOSTBCs achieving the upper bound of the data-rate is very
hard due to the fact that the covariance matrix of the noise term
at the destination terminal is non-diagonal in general. Therefore,
we restrict our interests to a subset of the DOSTBCs, whose codes
result in a diagonal noise covariance matrix at the destination
terminal. We refer to the codes in this subset as the
\emph{row-monomial DOSTBCs} and derive an upper bound of the
data-rate of the row-monomial DOSTBC. This upper bound is equal to
or slightly smaller than that of the DOSTBC; while it is much
higher than that of the repetition-based cooperative strategy.
Furthermore, we develop the systematic construction methods of the
row-monomial DOSTBCs achieving the upper bound of the data-rate.

The rest of this paper is organized as follows. Section
\ref{sec:sys} describes the cooperative network considered in this
paper. In Section \ref{sec:DOSTBC}, we first define the DOSTBCs
and then derive an upper bound of the data-rate of the DOSTBC. In
Section \ref{sec:DOSTBCw}, the row-monomial DOSTBCs are first
defined and an upper bound of the data-rate of the row-monomial
DOSTBC is then derived. Section \ref{sec:code} presents the
systematic construction methods of the DOSTBCs and row-monomial
DOSTBCs achieving the upper bound of the data-rate. We present
some numerical results in Section \ref{sec:num} to evaluate the
performance of the DOSTBCs and row-monomial DOSTBCs. The paper is
concluded in Section \ref{sec:conl}.

\emph{Notations:} Bold upper and lower letters denote matrices and
row vectors, respectively. Also, $\textrm{diag}[x_1,\cdots,x_K]$
denotes the $K\times K$ diagonal matrix with $x_1,\cdots,x_K$ on
its main diagonal; $\bmzero_{k_1\times k_2}$ the $k_1\times k_2$
all-zero matrix; $\bmI_{T\times T}$ the $T\times T$ identity
matrix; $\mathrm{det}(\cdot)$ the determinant of a matrix;
$[\cdot]_k$ the $k$-th entry of a vector; $[\cdot]_{k_1,k_2}$ the
$(k_1,k_2)$-th entry of a matrix; $(\cdot)^{*}$ the complex
conjugate; $(\cdot)^H$ the Hermitian; $(\cdot)^T$ the transpose.
For two real numbers $a$ and $b$, $\lceil a \rceil$ denotes the
ceiling function of $a$, i.e. the smallest integer bigger than
$a$; $\lfloor a\rfloor$ the floor function of $a$, i.e. the
largest integer smaller than $a$; $\mod(a,b)$ the modulo
operation, i.e. $\mod(a,b) = a - b\lfloor a/b \rfloor$. For two
sets ${\cal{S}}_1$ and ${\cal{S}}_2$, ${\cal{S}}_1-{\cal{S}}_2$
denotes the set whose elements are in ${\cal{S}}_1$ but not in
${\cal{S}}_2$.

%%%%%%%%%%%%%%%%%%%%%%%%%%%%%%%%%%%%%%%%%%%%%%%%%%%%%%%%%%%%%%%%%%%%

\section{System Model}\label{sec:sys}

Consider a cooperative network with one source terminal, $K$ relay
terminals, and one destination terminal. Every terminal has only
one antenna and is constrained to be half-duplex, i.e. a terminal
can not receive and transmit signals simultaneously. Denote the
channel from the source terminal to the $k$-th relay terminal by
$h_k$ and the channel from the $k$-th relay terminal to the
destination terminal by $f_k$. Both $h_k$ and $f_k$ are assumed to
be spatially uncorrelated complex Gaussian random variables with
zero mean and unit variance. We assume that the destination
terminal knows the instantaneous values of the channel
coefficients $h_k$ and $f_k$ by using training sequences; while
the source and relay terminals have no knowledge of the
instantaneous channel coefficients.

At the beginning, the source terminal transmits $N$ complex-valued
symbols over $N$ consecutive time slots.\footnote{If the
transmitted symbols are real-valued, it is easy to show that the
rate-one generalized real orthogonal design proposed in
\cite{tarokh} can be used in cooperative networks without any
changes, while achieving the single-symbol ML decodability and
full diversity order. Therefore, we focus on the complex-valued
symbols in this paper.} Let $\bms=[s_1,\cdots,s_N]$ denote the
symbol vector transmitted from the source terminal, where the
power of $s_n$ is $E_s$. Assume the coherence time of $h_k$ is
larger than $N$; then the received signal vector $\bmy_k$ at the
$k$-th relay terminal is given by
\begin{equation}
\bmy_k = h_k\bms+\bmn_k,
\end{equation}
where $\bmn_k = [n_{k,1},\cdots,n_{k,N}]$ is the additive noise at
the $k$-th relay terminal and is assumed to be uncorrelated
complex Gaussian with zero mean and identity covariance matrix. In
this paper, all the relay terminals are working in the
amplify-and-forward mode and the amplifying coefficient $\rho$ is
chosen to be $\sqrt{E_r/(1+E_s)}$ for every relay terminal, where
$E_r$ is the transmission power per use of every relay
terminal.\footnote{In many previous papers such as \cite{nabar},
\cite{jing1}, and \cite{jing2}, the same choice of
$\rho=\sqrt{E_r/(1+E_s)}$ has been made.} In order to construct a
distributed space-time code, every relay terminal multiplies
$\bmy_k$ and $\bmy_k^*$ with $\bmA_k$ and $\bmB_k$, respectively,
and then sum up these two products.\footnote{This construction
method originates from the construction of a linear space-time
code for co-located multiple-antenna systems, where the
transmitted signal vector from the $k$-th antenna is
$\bms\bmA_k+\bms^*\bmB_k$ \cite{wang}. Since we consider the
amplify-and-forward cooperative networks, the relay terminals do
not have the estimate of $\bms$. Therefore, they use $\bmy_k$ and
$\bmy_k^*$, which contain the information of $\bms$, to construct
the transmitted signal vector.} The dimension of $\bmA_k$ and
$\bmB_k$ is $N\times T$. Thus, the transmitted signal vector
$\bmx_k$ from the $k$-th relay terminal is given by
\begin{eqnarray}\label{eqn:bmxk}
\bmx_k &=& \rho(\bmy_k\bmA_k+\bmy_k^{*}\bmB_k)\nonumber\\
&=&\rho h_k \bms \bmA_k +\rho h_k^{*}
\bms^{*}\bmB_k+\rho\bmn_k\bmA_k+\rho\bmn_k^{*}\bmB_k.
\end{eqnarray}

Assume the coherence time of $f_k$ is larger than $T$; then the
received signal vector $\bmy_D$ at the destination terminal is
given by
\begin{eqnarray}
\bmy_D &=& \sum_{k=1}^K f_k\bmx_k + \bmn_D\nonumber\\
&=&\sum_{k=1}^K(\rho f_k h_k \bms \bmA_k +\rho f_k h_k^{*}
\bms^{*}\bmB_k)+\sum_{k=1}^K(\rho f_k\bmn_k\bmA_k+\rho
f_k\bmn_k^{*}\bmB_k)+\bmn_D, \label{eqn:ydfull2}
\end{eqnarray}
where $\bmn_D = [n_{D,1},\cdots,n_{D,T}]$ is the additive noise at
the destination terminal and is assumed to be uncorrelated complex
Gaussian with zero mean and identity covariance matrix. Define
$\bmw$, $\bmX$, and $\bmn$ as follows:
\begin{eqnarray}
\bmw&=&[\rho f_1,\cdots,\rho f_K]\\
\bmX&=&[h_1\bms\bmA_1+h_1^{*}\bms^{*}\bmB_1,\cdots,h_K\bms\bmA_K+h_K^{*}\bms^{*}\bmB_K]^{T}\label{eqn:X}\\
\bmn&=&\sum_{k=1}^K(\rho f_k\bmn_k\bmA_k+\rho
f_k\bmn_k^{*}\bmB_k)+\bmn_D;\label{eqn:n}
\end{eqnarray}
then we can rewrite (\ref{eqn:ydfull2}) in the following way
\begin{equation}
\bmy_D = \bmw \bmX +\bmn.\label{eqn:yd}
\end{equation}
Because the matrix $\bmX$ contains $N$ information-bearing
symbols, $s_1,\cdots,s_N$, and it lasts for $T$ time slots, the
data-rate of $\bmX$ is equal to $N/T$.\footnote{Considering the
$N$ time slots used by the source terminal to transmit the symbol
vector $\bms$, the data-rate of the entire transmission scheme is
$N/(N+T)$. In this paper, because we focus on the design of
$\bmX$, we will use the data-rate $N/T$ of $\bmX$ as the metric to
evaluate the bandwidth efficiency, as we have mentioned in Section
\ref{sec:intro}. Actually, once $N/T$ is known, it is very easy to
evaluate $N/(N+T)$.} From (\ref{eqn:n}), it is easy to see that
the mean of $\bmn$ is zero and the covariance matrix
$\bmR={\mathrm{E}}\left\{\bmn^H\bmn\right\}$ of $\bmn$ is given by
\begin{eqnarray}
\bmR%&=& {\mathrm{E}}\left\{\bmn^H\bmn\right\}\nonumber\\
%&=&\sum_{k=1}^K
%{\mathrm{E}}\left\{\left(\rho f_k\bmn_k\bmA_k+\rho
%f_k\bmn_k^{*}\bmB_k\right)^H\left(\rho f_k\bmn_k\bmA_k+\rho
%f_k\bmn_k^{*}\bmB_k\right)\right\}+{\mathrm{E}}\{\bmn_D^H\bmn_D\}\nonumber\\
&=& \sum_{k=1}^K \left(|\rho f_k|^2\left(\bmA_k^H\bmA_k
+\bmB_k^H\bmB_k\right)\right)+\bmI_{T\times T}.\label{eqn:R}
\end{eqnarray}
%In the following, we will use $R_{t_1,t_2}$ to denote the
%$(t_1,t_2)$-th entry of $\bmR$.

%%%%%%%%%%%%%%%%%%%%%%%%%%%%%%%%%%%%%%%%%%%%%%%%%%%%%%%%%%%%%%%%%%%%%

\section{Distributed Orthogonal Space-Time Block
Codes}\label{sec:DOSTBC}

In this section, we will define the DOSTBCs at first. Then, in
order to derive an upper bound of the data-rate of the DOSTBC,
some conditions on $\bmA_k$ and $\bmB_k$ are presented. Lastly,
based on those conditions, the upper bound is derived.

%\subsection{Definition of Distributed Orthogonal Space-Time Block Code}

From (\ref{eqn:yd}), the ML estimate $\hat{\bms}$ of $\bms$ is
given by
\begin{eqnarray}
\hat{\bms}&=& \arg\min_{\bmsspt\in {\cal{C}}} (\bmy_D
-\bmw\bmX)\bmR^{-1}(\bmy_D -\bmw\bmX)^H\nonumber\\
%&=& \arg\min_{\bms\in {\cal{\bmS}}} \left(||\bmy_D||^2
%-2\Re(\bmw\bmX\bmy_D^H) +||\bmw\bmX||^2 \right)\\
&=&\arg\min_{\bmsspt\in {\cal{C}}} \left(
-2\Re\left(\bmw\bmX\bmR^{-1}\bmy_D^H\right) +
\bmw\bmX\bmR^{-1}\bmX^H\bmw^H \right),\label{eqn:ML3}
\end{eqnarray}
where ${\cal{C}}$ is the set containing all the possible symbol
vector $\bms$ and it depends on the modulation scheme of $s_n$.
Inspired by the definition of the generalized orthogonal designs
\cite{tarokh,wang}, we define the DOSTBCs in the following way.

\vspace{3mm}

%\begin{defi}\label{defi:DOSTBC}
{\it Definition 1:} A $K\times T$ matrix $\bmX$ is called a
Distributed Orthogonal Space-Time Block Code (DOSTBC) in variables
$s_1,\cdots,s_N$ if the following two conditions are satisfied:
\begin{description}
    \item[D1.1)] The entries of $\bmX$ are 0, $\pm h_k s_n$,
    $\pm h_k^{*}s_n^{*}$, or multiples of these indeterminates by $\textbf{j}$, where $\textbf{j} =
    \sqrt{-1}$.
    \item[D1.2)] The matrix $\bmX$ satisfies the following equality
    \begin{eqnarray}\label{eqn:orth}
    \bmX\bmR^{-1}\bmX^H &=& |s_1|^2 \bmD_1 +\cdots +|s_N|^2
    \bmD_N,
    \end{eqnarray}
    where $\bmD_n$ is
    \begin{eqnarray}\label{eqn:Dn}
    \bmD_n&=&{\mathrm{diag}}[|h_1|^2D_{n,1},\cdots,|h_K|^2D_{n,K}]
    \end{eqnarray}
and $D_{n,1},\cdots,D_{n,K}$ are non-zero.\footnote{In the
definition of the generalized complex orthogonal designs,
$D_{n,1},\cdots,D_{n,K}$ are constrained to be strictly positive
in order to ensure that every row of $\bmX$ has at least one entry
containing $ s_n$ or $ s_n^*$ \cite{tarokh,wang}. However, for the
DOSTBCs, $D_{n,1},\cdots,D_{n,K}$ depend on $f_k$ through $\bmR$,
and hence, they are actually random variables. Therefore, we can
not constrain $D_{n,1},\cdots,D_{n,K}$ to be strictly positive.
Instead, we constrain them to be non-zero, which also ensures that
every row of $\bmX$ has at least one entry containing $ s_n$ or $
s_n^*$.}
\end{description}
%\end{defi}
\vspace{3mm}

Substituting (\ref{eqn:orth}) into (\ref{eqn:ML3}), it is easy to
show that the DOSTBCs are single-symbol ML decodable. Furthermore,
the numerical results given in Section \ref{sec:num} will
demonstrate that the DOSTBCs can achieve the full diversity order
$K$.\footnote{In \cite{jing1,jing2}, Jing \emph{et al.} considered
the cooperative networks using linear dispersion codes and they
analytically showed that the full diversity order $K$ could be
achieved. However, the authors constrained $\bmA_k\pm\bmB_k$ to be
orthogonal matrices and it greatly simplified the proof. In this
paper, in order to construct the codes achieving the upper bound
of the data-rate, we do not constrain $\bmA_k\pm\bmB_k$ to be
orthogonal. For example, in Subsection \ref{sub:2l+12m+1}, the
associated matrices $\bmA_k\pm\bmB_k$ of $\bmX(5,5)$, which
achieves the upper bound of the data-rate, are not orthogonal.
Therefore, it will be very hard to analytically prove the full
diversity order of the DOSTBCs. Instead, we present some numerical
results in Section \ref{sec:num} to show that the DOSTBCs achieve
the full diversity order $K$ indeed. Some intuitive explanations
are also provided after Theorem \ref{thm:necessary}.} Therefore,
compared with the repetition-based cooperative strategy, the
DOSTBCs have the same decoding complexity and diversity order. In
this paper, we will show that the DOSTBCs have much better
bandwidth efficiency than the repetition-based cooperative
strategy.

In order to show the higher bandwidth efficiency of the DOSTBCs,
it is desirable to derive an upper bound of the data-rate of the
DOSTBC. Generally, it is very hard to derive the upper
bound by directly using the conditions on $\bmX$ in Definition 1%\ref{defi:DOSTBC}
. From (\ref{eqn:X}), we note that the structure of $\bmX$ is
purely decided by $\bmA_k$ and $\bmB_k$. Therefore, we transfer
the conditions on $\bmX$ into some conditions on $\bmA_k$ and
$\bmB_k$. We first present some fundamental conditions on $\bmA_k$
and $\bmB_k$ from D1.1, which will be used throughout this paper.
For convenience, we define that a matrix is said to be {\it
column-monomial (row-monomial)} if there is at most one non-zero
entry on every column (row) of it.
\begin{lem}\label{thm:propertyAB}
If a DOSTBC $\bmX$ in variables $s_1,\cdots,s_N$ exists, its
associated matrices $\bmA_k$ and $\bmB_k$, $1\leq k\leq K$,
satisfy the following conditions:
\begin{enumerate}
    \item The entries of $\bmA_k$ and $\bmB_k$ can only be 0,
    $\pm 1$, or $\pm \textbf{j}$.
    \item $\bmA_k$ and $\bmB_k$ can not have non-zero entries at
    the same position.
    \item $\bmA_k$, $\bmB_k$, and $\bmA_k+\bmB_k$ are column-monomial.
    \end{enumerate}
\end{lem}
\begin{proof}
See Appendix A.
\end{proof}
Secondly, we derive some conditions on $\bmA_k$ and $\bmB_k$ that
are equivalent with the orthogonal condition (\ref{eqn:orth}) on
$\bmX$.
\begin{lem}\label{thm:originalcond}
The orthogonal condition (\ref{eqn:orth}) on $\bmX$ holds if and
only if
\begin{eqnarray}
\bmA_{k_1}\bmR^{-1}\bmA_{k_2}^H &=& \bmzero_{N\times N},
\hspace{2cm} 1 \leq k_1 \neq k_2 \leq K\label{eqn:cond1}
\\
\bmB_{k_1}\bmR^{-1}\bmB_{k_2}^H &=& \bmzero_{N\times N},
\hspace{2cm} 1 \leq k_1
\neq k_2 \leq K\label{eqn:cond2}\\
\bmA_{k_1} \bmR^{-1} \bmB_{k_2}^H + \bmB_{k_2}^{*}\bmR^{-1}
\bmA_{k_1}^T &=& \bmzero_{N\times N}, \hspace{2cm} 1 \leq k_1,k_2 \leq K\label{eqn:cond3}\\
\bmB_{k_1}\bmR^{-1}\bmA_{k_2}^H +
\bmA_{k_2}^*\bmR^{-1}\bmB_{k_1}^T
&=& \bmzero_{N\times N}, \hspace{2cm} 1 \leq k_1,k_2 \leq K\label{eqn:cond4}\\
\bmA_k \bmR^{-1} \bmA_k^H + \bmB^{*}_k \bmR^{-1} \bmB^T_ k&=&
{\mathrm{diag}}[D_{1,k},\cdots,D_{N,k}], \hspace{2cm} 1 \leq k
\leq K.\label{eqn:cond5}
\end{eqnarray}
\end{lem}
\begin{proof}
See Appendix B.
\end{proof}

One possible way to derive the upper bound of the data-rate of the
DOSTBC is by using the conditions
(\ref{eqn:cond1})--(\ref{eqn:cond5}) in Lemma
\ref{thm:originalcond}. However, the existence of $\bmR^{-1}$ in
those conditions make the derivation very hard. Therefore, we
simplify the conditions (\ref{eqn:cond1})--(\ref{eqn:cond5}) in
the following theorem by eliminating $\bmR^{-1}$.

%Furthermore, similar to the definition of generalized complex
%orthogonal design, we constrain that $\bmX$ does not contain the
%linear combination of $\pm h_k s_n$ and $\pm h_k^{*} s_n^{*}$.

\begin{thm}\label{thm:necessary}
If a DOSTBC $\bmX$ in variables $s_1,\cdots,s_N$ exists, we have
\begin{eqnarray}\label{eqn:orthE}
\bmX\bmX^H&=&|s_1|^2\bmE_1 + \cdots+ |s_N|^2\bmE_N,
\end{eqnarray}
where $\bmE_n$ is
\begin{eqnarray}\label{eqn:En}
\bmE_n &=& {\mathrm{diag}}[|h_1|^2 E_{n,1},\cdots,|h_K|^2 E_{n,K}]
\end{eqnarray}
and $E_{n,1},\cdots,E_{n,K}$ are strictly positive.
\end{thm}
\begin{proof}
See Appendix C.
\end{proof}

Theorem \ref{thm:necessary} can provide some intuitive
explanations on the full diversity of the DOSTBCs. According to
Theorem \ref{thm:necessary}, $E_{n,1},\cdots,E_{n,K}$, $1\leq n
\leq N$, are strictly positive for a DOSTBC $\bmX$. This implies
that every row of $\bmX$ has at least one entry containing $s_n$,
or equivalently, every relay terminal transmits at least one
symbol containing $s_n$, $1\leq n \leq N$. Therefore, the
destination terminal receives $K$ replica of $s_n$ through $K$
independent fading channels and the diversity order of the DOSTBCs
should be $K$.

Furthermore, the new orthogonal condition (\ref{eqn:orthE}) on
$\bmX$ and its equivalent conditions
(\ref{eqn:Econd1})--(\ref{eqn:Econd5}) on $\bmA_k$ and $\bmB_k$
are much simpler than the original orthogonal condition
(\ref{eqn:orth}) and its equivalent conditions
(\ref{eqn:cond1})--(\ref{eqn:cond5}). The new conditions enable us
to find an upper bound of the data-rate of the DOSTBC.
\begin{thm}\label{thm:upperDOSTBC}
If a DOSTBC $\bmX$ in variables $s_1,\cdots,s_N$ exists, its
data-rate $\mathrm{Rate}$ satisfies the following inequality:
\begin{equation}\label{eqn:rate}
{\mathrm{Rate}}=\frac{N}{T} \leq \frac{N}{\lceil \frac{NK}{2}
\rceil}.
\end{equation}
%where $\lceil (NK)/2 \rceil$ denotes the ceiling function of
%$(NK)/2 $.
\end{thm}
\begin{proof}
See Appendix D.
\end{proof}

Compared with the data-rate $1/K$ of the repetition-based
cooperative strategy, it is easy to see that the upper bound of
(\ref{eqn:rate}) is approximately twice higher, which implies that
the DOSTBCs potentially have much higher bandwidth efficiency over
the repetition-based cooperative strategy. Furthermore, it is
worthy of addressing that the DOSTBCs have the same decoding
complexity and diversity order as the repetition-based cooperative
strategy.

After obtaining the upper bound of the data-rate of the DOSTBC, a
natural question is how to construct the DOSTBCs achieving the
upper bound of (\ref{eqn:rate}). Unfortunately, only for the case
that $N$ and $K$ are both even, we can find such DOSTBCs and they
are given in Subsection \ref{subsec:2l2m}. For the other cases,
where $N$ or/and $K$ is odd, we could not find any DOSTBCs
achieving the upper bound of (\ref{eqn:rate}).\footnote{Actually,
we do not know if the upper bound of (\ref{eqn:rate}) is
achievable or not for these cases. Our conjecture is that the
upper bound of (\ref{eqn:rate}) can be tightened for these cases
and it should be the same as that of the row-monomial DOSTBC
defined in the next Section. Analytical proof has not been found
yet; but some intuitive explanations are provided in the last
paragraph of the next section.} We note that the major hindrance
comes from the fact that the noise covariance matrix $\bmR$ in
(\ref{eqn:R}) is not diagonal in general. In the next section,
thus, we will consider a subset of the DOSTBCs, whose codes result
in a diagonal $\bmR$.

%Therefore, we do some further investigations on row-monomial DOSTBCs,
%which result in white noise at the destination terminal. We show
%that the upper bound of the data-rate of row-monomial DOSTBCs is equal to
%or slightly smaller than that of DOSTBC. However, for any values
%of $N$ and $K$, systematic construction methods of the
%row-monomial DOSTBCs achieving the upper bound of the data-rate are given in Section \ref{sec:code}.

%%%%%%%%%%%%%%%%%%%%%%%%%%%%%%%%%%%%%%%%%%%%%%%%%%%%%%%%%%%%%%%%%%%%%55

\section{Row-Monomial Distributed Orthogonal Space-Time Block
Codes}\label{sec:DOSTBCw}

In this section, we first show that, if $\bmA_k$ and $\bmB_k$ are
row-monomial, the covariance matrix $\bmR$ becomes diagonal. Then
we define a subset of the DOSTBCs, whose associated matrices
$\bmA_k$ and $\bmB_k$ are row-monomial, and hence, we refer to the
codes in this subset as the row-monomial DOSTBCs. Lastly, an upper
bound of the data-rate of the row-monomial DOSTBC is derived.

As we stated in Section \ref{sec:DOSTBC}, the non-diagonality of
$\bmR$ makes the construction of the DOSTBCs achieving the
upper-bound of (\ref{eqn:rate}) very hard. Thus, we restrict our
interests to a special subset of the DOSTBCs, where $\bmR$ is
diagonal. In the following, we show that the diagonality of $\bmR$
is equivalent with the row-monomial condition of $\bmA_k$ and
$\bmB_k$.

\begin{thm}\label{thm:whitenoise}
The matrix $\bmR$ in (\ref{eqn:R}) is a diagonal matrix if and
only if $\bmA_k$ and $\bmB_k$ are row-monomial.
\end{thm}
\begin{proof}
See Appendix E.
\end{proof}

Based on Theorem \ref{thm:whitenoise}, we define the row-monomial
DOSTBCs in the following way.

\vspace{3mm}
%\begin{defi}\label{defi:DOSTBCw}
{\it Definition 2:} A $K\times T$ matrix $\bmX$ is called a
row-monomial DOSTBC in variables $s_1,\cdots,s_N$ if it satisfies
D1.1
and D1.2 in Definition 1 %\ref{defi:DOSTBC}
 and its associated matrices $\bmA_k$ and $\bmB_k$, $1\leq k \leq K$,
 are both
row-monomial.
%\begin{description}
%    \item[D6.1:] Every entry of $\bmX$ is 0, $\pm h_k s_n$,
%    $\pm h_k^{*}s_n^{*}$, or multiples of them by $\textbf{j}$, where $\textbf{j} =
%    \sqrt{-1}$.
%    \item[D6.2:] $\bmX\bmR^{-1}\bmX^H = |s_1|^2 \bmD_1 +\cdots +|s_N|^2
%    \bmD_N$, where $\bmR$ and $\bmD_n$ are defined by
%    (\ref{eqn:R}) and (\ref{eqn:Dn}), respectively.
%    \item[D6.3:] $\bmA_k$ and $\bmB_k$ are row-monomial.
%    The noise term $\bmn$ at the destination terminal is
%    temporally white, i.e. $\bmR$ is a diagonal matrix.
%\end{description}
%\end{defi}
\vspace{3mm}

Obviously, the row-monomial DOSTBCs are single-symbol ML decodable
because they are in a subset of the DOSTBCs. Numerical results in
Section \ref{sec:num} show that they also achieve the full
diversity order $K$. Furthermore, all the results in Section
\ref{sec:DOSTBC} are still valid for the row-monomial DOSTBCs. In
particular, the upper-bound given by (\ref{eqn:rate}) can still
serve as an upper-bound of the data-rate of the row-monomial
DOSTBC. However, due to some special properties of the
row-monomial DOSTBCs, a tighter upper bound can be derived. To
this end, we need to present several conditions on $\bmA_k$ and
$\bmB_k$ at first. In this paper, two matrices $\bmA$ and $\bmB$
are said to be {\it column-disjoint}, if $\bmA$ and $\bmB$ can not
contain non-zero entries on the same column simultaneously, i.e.
if a column in $\bmA$ contains a non-zero entry at any row, then
all the entries of the same column in $\bmB$ must be zero;
conversely, if a column in $\bmB$ contains a non-zero entry at any
row, then all the entries of the same column in $\bmA$ must be
zero.
\begin{lem}\label{thm:rowmonomial}
If a row-monomial DOSTBC $\bmX$ in variables $s_1,\cdots,s_N$
exists, its associated matrices $\bmA_k$ and $\bmB_k$, $1\leq k
\leq K$, satisfy the following two conditions:

1) $\bmA_{k_1}$ and $\bmA_{k_2}$ are column-disjoint for $k_1 \neq
k_2$.

2) $\bmB_{k_1}$ and $\bmB_{k_2}$ are column-disjoint for $k_1 \neq
k_2$.
%\begin{enumerate}
%    \item $\bmA_k$ and $\bmB_k$ are row-monomial.
%    \item When $k_1 \neq k_2$, $\bmA_{k_1}$ and $\bmA_{k_2}$ are
%     column-disjoint; $\bmB_{k_1}$ and $\bmB_{k_2}$ are
%     column-disjoint.
%\end{enumerate}
\end{lem}
\begin{proof}
See Appendix F.
\end{proof}

%Intuitively, Lemma \ref{thm:rowmonomial} means that $\pm h_k
%s_n$ and $\pm h_k^* s_n^*$ appear in the $k$-th row of $\bmX$ for
%at most one time. Otherwise, the noise term $\bmn$ will be
%correlated. But, note that the $k$-th row of $\bmX$ can contain
%both $\pm h_k s_n$ and $\pm h_k^* s_n^*$, which will not make
%$\bmn$ correlated.

Lemma \ref{thm:rowmonomial} is crucial to find the upper
bound of the data-rate of the row-monomial DOSTBC. According to Definition 2%\ref{defi:DOSTBCw}
, if $\bmX$ is a row-monomial DOSTBC, there are two types of
non-zero entries in it: 1) the entries containing $\pm h_k s_n$ or
the multiples of it by $\textbf{j}$; 2) the entries containing
$\pm h_k^* s_n^*$ or the multiples of it by $\textbf{j}$. In the
following, we will refer to the first type of entries as the {\it
non-conjugate} entries and refer to the second type of entries as
the {\it conjugate} entries. Lemma \ref{thm:rowmonomial} implies
that any column in $\bmX$ can not contain more than one
non-conjugate entry or more than one conjugate entries. However,
one column in $\bmX$ can contain one non-conjugate entry and one
conjugate entry at the same time. Therefore, the columns in $\bmX$
can be partitioned into two types: 1) the columns containing one
non-conjugate entry or one conjugate entry; 2) the columns
containing one non-conjugate entry and one conjugate entry. In the
following, we will refer to the first type of columns as the
Type-I columns and refer to the second type of columns as the
Type-II columns. For the Type-II columns, we have the following
lemma.
\begin{lem}\label{thm:tpyeii}
If a row-monomial DOSTBC $\bmX$ in variables $s_1,\cdots,s_N$
exists, the Type-II columns in $\bmX$ have the following
properties:
\begin{enumerate}
    \item The total number of the Type-II columns in $\bmX$ is even.
    \item In all the Type-II columns of $\bmX$, the total
number of the entries containing $s_n$ or $s_n^*$, $1\leq n \leq
N$, is even.
\end{enumerate}
\end{lem}
\begin{proof}
See Appendix G.
\end{proof}

Since the data-rate of $\bmX$ is defined as $N/T$, improving the
data-rate of $\bmX$ is equivalent to reducing the length $T$ of
$\bmX$, when $N$ is fixed. Furthermore, we note that a Type-II
column contains two non-zero entries; while a Type-I column
contains only one non-zero entries. Therefore, if all the non-zero
entries in $\bmX$ are contained in the Type-II columns, the
data-rate of $\bmX$ achieves the maximum value. Unfortunately, in
some circumstances, not all the non-zero entries in $\bmX$ can be
contained in the Type-II columns. In those circumstances, in order
to reduce $T$, we need to make $\bmX$ contain non-zero entries in
the Type-II columns as many as possible. Based on this and Lemmas
\ref{thm:rowmonomial} and \ref{thm:tpyeii}, we derive an upper
bound of the data-rate of the row-monomial DOSTBC and the result
is given in the following theorem.

\begin{thm}\label{thm:upperDOSTBCw}
If a row-monomial DOSTBC $\bmX$ in variables $s_1,\cdots,s_N$
exists, its data-rate ${\mathrm{Rate}}_r$ satisfies the following
inequality:
\begin{equation}\label{eqn:upperDOSTBCw}
{\mathrm{Rate}}_r =\frac{N}{T} \leq \left\{
\begin{array}{cl}
  \frac{1}{m}, & {\mathrm{when}} \hspace{3mm} N=2l, K=2m \\
  \frac{2l+1}{2lm+2m}, & {\mathrm{when}} \hspace{3mm} N=2l+1, K=2m  \\
  \frac{1}{m+1}, & {\mathrm{when}} \hspace{3mm} N=2l, K=2m+1  \\
  \min\left(\frac{2l+1}{2lm+2m+l+1}, \frac{2l+1}{2lm+2l+m+1}\right), & {\mathrm{when}} \hspace{3mm} N=2l+1, K=2m+1  \\
\end{array}%
\right.,
\end{equation}
where $l$ and $m$ are positive integers.
\end{thm}
%\vspace{5mm}
\begin{proof}
See Appendix H.
\end{proof}

\begin{figure}
\begin{center}
\subpostscript{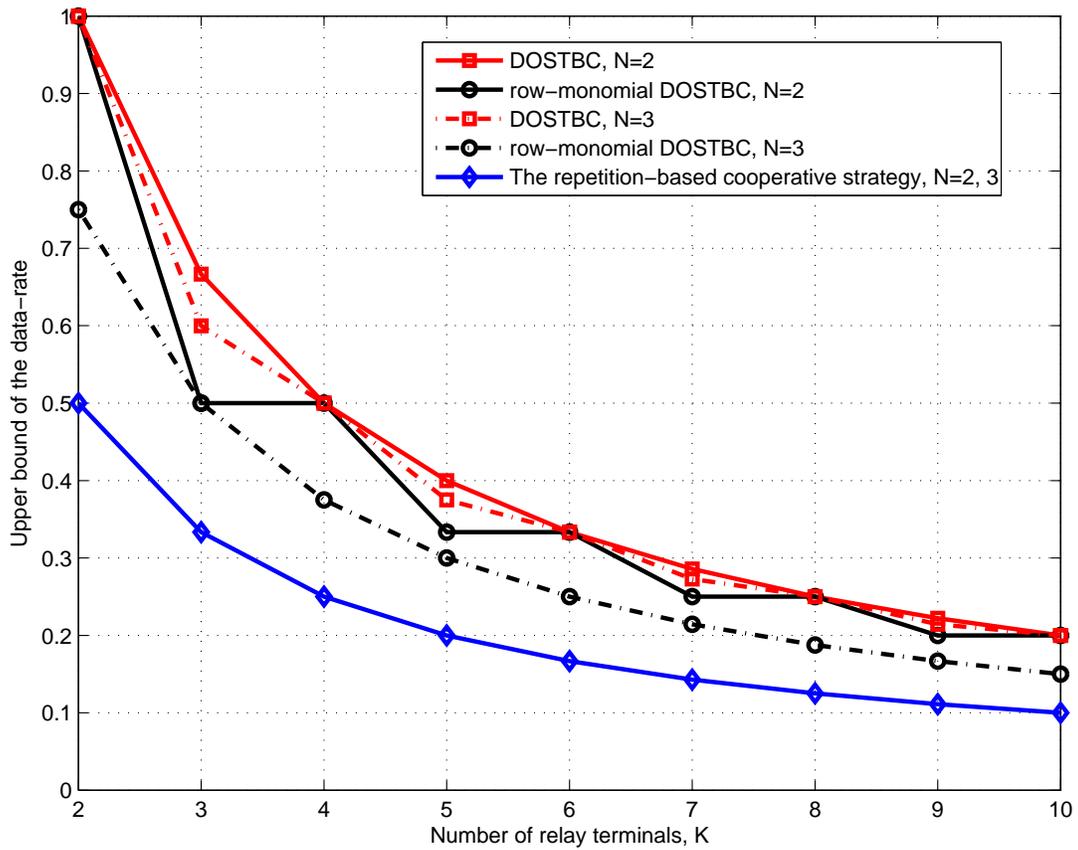}{0.9\textwidth}
\end{center}
\caption{Comparison of the upper bounds of the data-rates of the
DOSTBC, row-monomial DOSTBC, and repetition-based cooperative
strategy, $N=2$, $3$.} \label{fig:rateN2N3}
\end{figure}

According to (\ref{eqn:upperDOSTBCw}), the row-monomial DOSTBCs
have much better bandwidth efficiency than the repetition-based
cooperative strategy. In order to compare the bandwidth efficiency
of the DOSTBCs and row-monomial DOSTBCs, we summarize the upper
bounds given
by (\ref{eqn:rate}) and (\ref{eqn:upperDOSTBCw}) in Fig. \ref{fig:rateN2N3} and Table I. %\ref{tab:bounds}
For the first case that $N$ and $K$ are even, the upper bounds of
the data-rates of the DOSTBC and row-monomial DOSTBC are both
$1/m$, i.e. they have the same bandwidth efficiency. A systematic
method to construct the DOSTBCs and row-monomial DOSTBCs achieving
the upper bound $1/m$ is given in Subsection \ref{subsec:2l2m}.
For the other three cases, the upper bound of the data-rate of the
row-monomial DOSTBC is smaller than that of the DOSTBC. However,
the difference is very marginal and it goes to zero asymptotically
when $K$ goes to infinite. Therefore, when a large number of relay
terminals participate in the cooperation, the row-monomial DOSTBCs
will have almost the same bandwidth efficiency as the DOSTBCs.
Furthermore, systematic construction methods of the row-monomial
DOSTBCs achieving the upper bound of (\ref{eqn:upperDOSTBCw}) are
given in Subsections \ref{sub:2l+12m}, \ref{sub:2l2m+1} and
\ref{sub:2l+12m+1} for these three cases.

Actually, we conjecture that the upper bound of (\ref{eqn:rate})
can be tightened for those three cases and it should be the same
as (\ref{eqn:upperDOSTBCw}) irrespective of the values of $N$ and
$K$. The proof has not been found yet; but we provide some
intuitive reasons in the following. As we mentioned before, any
column in a row-monomial DOSTBC can not contain more than two
non-zero entries due to the row-monomial condition of $\bmA_k$ and
$\bmB_k$. On the other hand, one column in a DOSTBC can contain at
most $K$ non-zero entries. However, including more than two
non-zero entries in one column may be harmful to the maximum
data-rate. Let us consider the following example. Assume the first
column of $\bmX$ contains three non-zero entries at the first,
second, and third row. Without loss of generality, those non-zero
entries are assumed to be $h_1s_{n_1}$, $h_2^*s_{n_2}^*$, and
$h_3s_{n_3}$. Thus, in order to make the first and third row
orthogonal with each other, there must be another column
containing $-h_1s_{n_1}$ and $h_3s_{n_3}$ at the first and third
row, respectively. Therefore, there are two non-zero entries at
the first row of $\bmX$ containing $s_{n_1}$ and it is detrimental
to the data-rate of $\bmX$.\footnote{By comparison, we find that
(\ref{eqn:orthE}) and (\ref{eqn:En}) are similar with the
definition of the generalized orthogonal designs
\cite{tarokh,wang}. Actually, for any realizations of the channel
coefficients $h_1,\cdots,h_K$, (\ref{eqn:orthE}) and
(\ref{eqn:En}) become exactly the same as the definition of the
generalized orthogonal designs. It implies that theorems from the
generalized orthogonal designs should be still valid for the
DOSTBCs. For the generalized orthogonal designs, including the
symbol $s_n$ for only once in every row is enough to make the
codes have the full diversity order, where the symbol $s_n$ can
appear as either $s_n$ or $s_n^*$. Sometimes, in order to increase
the data-rate, one row may include $s_n$ and $s_n^*$ at the same
time. However, after searching the existing generalized orthogonal
designs achieving the highest data-rate, we can not find any code
containing more than one $s_n$ or $s_n^*$ in one row
\cite{tarokh}, \cite{su1}--\cite{liang1}. Therefore, we conjecture
that including more than one $s_n$ or $s_n^*$ in one row is
harmful to the data-rate of the generalized orthogonal designs.
Due to the similarity between the DOSTBCs and generalized
orthogonal designs, this conjecture should still hold for the
DOSTBCs.} As in this example, including three non-zero entries in
one column will make two non-zero entries at one row contain the
same symbol, and hence, it will decrease the data-rate. The same
argument can be made when more than three non-zero entries are
included in one column. Therefore, we believe that including more
than two non-zero entries in one column will incur a loss of the
data-rate of the code. Since the row-monomial DOSTBCs never
contain more than two non-zero entries in one column, we
conjecture that the upper bound of the data-rate of the
row-monomial DOSTBC should be the upper bound of the data-rate of
the DOSTBC as well.

%%%%%%%%%%%%%%%%%%%%%%%%%%%%%%%%%%%%%%%%%%%%%%%%%%%%%%%%%%%%%%%%%%%%

\section{Systematic Construction of the DOSTBCs and
Row-Monomial DOSTBCs Achieving the Upper Bound of the
Data-Rate}\label{sec:code}

In this section, we present the systematic construction methods of
the DOSTBCs and row-monomial DOSTBCs achieving the upper bound of
the data-rate. For given $N$ and $K$, we use $\bmX(N,K)$ to denote
the codes achieving the upper bound of the data-rate. There are
four different cases depending on the values of $N$ and $K$.

\subsection{$N=2l$ and $K=2m$}\label{subsec:2l2m}

For this case, the systematic construction method of the DOSTBCs
achieving the upper bound of (\ref{eqn:rate}) is found. For
convenience, we will use $\bmA_k(:,t_1:t_2)$ to denote the
submatrix consisting of the $t_1$-th, $t_1+1$-th, $\cdots$,
$t_2$-th columns of $\bmA_k$. Similarly, $\bmB_k(:,t_1: t_2)$
denotes the submatrix consisting of the $t_1$-th, $t_1+1$-th,
$\cdots$, $t_2$-th columns of $\bmB_k$. Furthermore, define
$\bmG_s$ as follows:
\begin{eqnarray}
\bmG_s&=&\left[%
\begin{array}{cc}
  0 & 1 \\
  1 & 0 \\
\end{array}%
\right].
\end{eqnarray}
Based on $\bmG_s$, two matrices $\bmG_A$ and $\bmG_B$ with
dimension $N\times N$ are defined:
\begin{eqnarray}
\bmG_A &=&{\mathrm{diag}}[1,-1,1,-1,\cdots,1,-1]\\
\bmG_B &=&{\mathrm{diag}}\left[\bmG_s,\cdots,\bmG_s\right].
%\left[%
%\begin{array}{ccccccc}
%  0 & 1  & 0 & 0  & \cdots  &   0 & 0 \\
%  1 & 0  & 0 & 0  & \cdots  &   0 & 0  \\
%  0 & 0  & 0 & 1  & \cdots  &   0 & 0 \\
%  0 & 0  & 1 & 0  & \cdots  &   0 & 0   \\
%  \vdots&\vdots&\vdots&\vdots&\ddots&\vdots&\vdots  \\
%   % &    &   &    &   &\ddots &   &  \\
%  0 & 0  & 0 & 0  & \cdots  &   0 & 1 \\
%  0 & 0  & 0 & 0  & \cdots  &   1 & 0 \\
%\end{array}%
%\right].
\end{eqnarray}
The proposed systematic construction method of the DOSTBCs
achieving the upper bound of (\ref{eqn:rate}) is as
follows:\footnote{It is easy to see that the codes generated by
this construction method are actually row-monomial DOSTBCs and
they also achieve the upper bound of (\ref{eqn:upperDOSTBCw}).}

Construction I:

{\it Initialization:} Set $p=1$. Set
$\bmA_k=\bmB_k=\bmzero_{N\times \infty}$, $1\leq k\leq K$, where
$\infty$ means that the length of the matrices is not decided yet.
%,where
%$\bmzero_{N\times 2lm}$ is a $N\times 2lm$ with all zero entries.

{\it Step 1:} Set $\bmA_{2p-1}(:,(p-1)N+1:pN) = \bmG_A$ and
$\bmB_{2p}(:,(p-1)N+1:pN) = \bmG_B$.

{\it Step 2:} Set $p=p+1$. If $p \leq m$, go to {\it Step 1};
otherwise, go to {\it Step 3}.

{\it Step 3:} Discard the all-zero columns at the tail of
$\bmA_{K-1}$ and $\bmB_K$. Set the length of $\bmA_k$ and
$\bmB_k$, $1\leq k\leq K$, equal to that of $\bmA_{K-1}$ and
$\bmB_K$.

{\it Step 4:} Calculate $\bmX(N,K)$ through (\ref{eqn:X}) by using
the matrices $\bmA_k$ and $\bmB_k$ obtained in {\it Steps 1--3},
and end the construction.

%As examples, we present two codes which are obtained by using the
%proposed construction method. When $K=2$ and $N=2$, the code is given by
%\begin{eqnarray}
%\bmX &=&\left[%
%\begin{array}{cc}
%  h_1 s_1 & -h_1 s_2 \\
%  h_2^*s_2^* & h_2^*s_1^* \\
%\end{array}%
%\right],
%\end{eqnarray}
The following lemma shows that Construction I generates the
DOSTBCs achieving the upper bound of (\ref{eqn:rate}) for any even
$N$ and $K$.
\begin{lem}
For any even $N=2l$ and $K=2m$, the codes generated by
Construction I achieve the data-rate $1/m$.
\end{lem}
\begin{proof}
In Construction I, the length of $\bmA_k$ and $\bmB_k$, $1\leq
k\leq K $, is decided by the length of $\bmA_{K-1}$ and $\bmB_K$.
Because $\bmA_{K-1}(:,(m-1)N+1:mN)$ and $\bmB_{K}(:,(m-1)N+1:mN)$
are set to be $\bmG_A$ and $\bmG_B$, respectively, when $p=m$, the
length of $\bmA_{K-1}$ and $\bmB_K$ is $mN$. Consequently, the
length of $\bmA_k$ and $\bmB_k$, $1\leq k\leq K $ is $mN$. By
(\ref{eqn:X}), the length of $\bmX(N,K)$ is $T$ and it is the same
as that of $\bmA_k$ and $\bmB_k$. Therefore, the value of $T$ is
$mN$, and hence, the data-rate of $\bmX(N,K)$ is $1/m$.
\end{proof}

For example, when $N=4$ and $K=4$, the code constructed by
Construction I is given by
\begin{eqnarray}
\bmX(4,4)&=& \left[%
\begin{array}{cccccccc}
  h_1s_1 & -h_1s_2 & h_1s_3 & -h_1s_4 & 0 & 0 & 0 & 0 \\
  h_2^*s_2^* & h_2^*s_1^* & h_2^*s_4^* & h_2s_3^* & 0 & 0 & 0 & 0 \\
  0 & 0 & 0 & 0 & h_3s_1 & -h_3s_2 & h_3s_3 & -h_3s_4 \\
  0 & 0 & 0 & 0 & h_4^*s_2^* & h_4^*s_1^* & h_4^*s_4^* & h_4s_3^* \\
\end{array}%
\right],
\end{eqnarray}
and it achieves the upper bound of the data-rate $1/2$.

\subsection{$N=2l+1$ and $K=2m$}\label{sub:2l+12m}

This case is equivalent with the case that $N=2l$ and $K=2m$ if
$s_N$ is not considered. Based on this, the proposed systematic
construction method of the row-monomial DOSTBCs achieving the
upper bound of (\ref{eqn:upperDOSTBCw}) is as follows:

Construction II:

{\it Step 1:} Neglect $s_N$ and construct a $K\times 2lm$ matrix
$\bmX_1$ in variables $s_1,\cdots,s_{N-1}$ by Construction I.

{\it Step 2:} Form a $K\times K$ diagonal matrix $\bmX_2 =
{\mathrm{diag}}[h_1s_N,\cdots,h_Ks_N]$.

{\it Step 3:} Let $\bmX(N,K) = [\bmX_1,\bmX_2]$ and end the
construction.

%As examples, we present two codes which are obtained by using the
%proposed construction method. When $K=2$ and $N=3$, the code is
%given by
%\begin{eqnarray}
%\bmX &=&\left[%
%\begin{array}{cccc}
%  h_1 s_1 & -h_1 s_2 &h_1s_3 & 0 \\
%  h_2^*s_2^* & h_2^*s_1^* &0 &h_1s_3\\
%\end{array}%
%\right],
%\end{eqnarray}
Because the length of $\bmX_1$ and $\bmX_2$ is $2lm$ and $K$,
respectively, the length of $\bmX(N,K)$ is $2lm+K$. Thus, the
data-rate of $\bmX(N,K)$ is $(2l+1)/(2lm+K)$, which is exactly the
same as the upper-bound of (\ref{eqn:upperDOSTBCw}).

For example, when $N=5$ and $K=4$, the code constructed by
Construction II is given by
\begin{equation}
\bmX(5,4)=\left[%
\begin{array}{ccccccccccccc}
  h_1s_1 & -h_1s_2 & h_1s_3 & -h_1s_4 & 0 & 0 & 0 & 0   &\vline& h_1s_5&0&0&0\\
  h_2^*s_2^* & h_2^*s_1^* & h_2^*s_4^* & h_2s_3^* & 0 & 0 & 0 & 0 &\vline&0&h_2s_5&0&0\\
  0 & 0 & 0 & 0 & h_3s_1 & -h_3s_2 & h_3s_3 & -h_3s_4&\vline& 0&0&h_3s_5&0 \\
  0 & 0 & 0 & 0 & h_4^*s_2^* & h_4^*s_1^* & h_4^*s_4^* & h_4s_3^*&\vline& 0&0&0&h_4s_5 \\
\end{array}%
\right],
\end{equation}
where the solid line illustrates the construction steps. The code
$\bmX(5,4)$ achieves the upper bound of the data-rate $5/12$.

For this case, the DOSTBCs achieving the upper bound of
(\ref{eqn:rate}) are not found.

\subsection{$N=2l$ and $K=2m+1$}\label{sub:2l2m+1}

This case is equivalent with the case that $N=2l$ and $K=2m$ if
the $K$-th relay terminal is not considered. Based on this, the
proposed systematic construction method of the row-monomial
DOSTBCs achieving the upper bound of (\ref{eqn:upperDOSTBCw}) is
as follows:

Construction III:

{\it Step 1:} Neglect the $K$-th relay terminal and construct a
$2m \times 2lm$ matrix $\bmX_1$ by Construction I.

{\it Step 2:} Form a vector $\bmx_2 = [h_K s_1, \cdots, h_K s_N]$

{\it Step 3:} Build a block diagonal matrix $\bmX(N,K) =
{\mathrm{diag}}[\bmX_1,\bmx_2]$ and end the construction.

%As examples, we present two codes which are obtained by using the
%proposed construction method. When $K=3$ and $N=2$, the code is
%given by
%\begin{eqnarray}
%\bmX &=&\left[%
%\begin{array}{cccc}
%  h_1 s_1 & -h_1 s_2 &0&0 \\
%  h_2^*s_2^* & h_2^*s_1^* &0&0\\
%  0&0& h_3s_1 &h_3 s_2\\
%\end{array}%
%\right],
%\end{eqnarray}
Because the length of $\bmX_1$ and $\bmx_2$ is $2lm$ and $N$,
respectively, the length of $\bmX(N,K)$ is $2lm+N$. Thus, the
data-rate of $\bmX(N,K)$ is $1/(1+m)$, which is exactly the same
as the upper-bound of (\ref{eqn:upperDOSTBCw}).

For example, when $N=4$ and $K=5$, the code constructed by
Construction III is given by
\begin{equation}
\bmX(4,5)= \left[%
\begin{array}{ccccccccccccc}
  h_1s_1 & -h_1s_2 & h_1s_3 & -h_1s_4 & 0 & 0 & 0 & 0 &\vline & 0 & 0 & 0 & 0\\
  h_2^*s_2^* & h_2^*s_1^* & h_2^*s_4^* & h_2s_3^* & 0 & 0 & 0 & 0 &\vline& 0 & 0 & 0 & 0\\
  0 & 0 & 0 & 0 & h_3s_1 & -h_3s_2 & h_3s_3 & -h_3s_4&\vline & 0 & 0 & 0 & 0\\
  0 & 0 & 0 & 0 & h_4^*s_2^* & h_4^*s_1^* & h_4^*s_4^* & h_4s_3^*&\vline & 0 & 0 & 0 & 0\\
\hline
  0 & 0 & 0 & 0 & 0 & 0 & 0 & 0 &\vline& h_5s_1 & h_5s_2 & h_5s_3 &
  h_5s_4\\
\end{array}%
\right],
\end{equation}
where the solid lines illustrate the construction steps. The code
$\bmX(4,5)$ achieves the upper bound of the data-rate $1/3$.

For this case, the DOSTBCs achieving the upper bound of
(\ref{eqn:rate}) are not found.

\subsection{$N=2l+1$ and $K=2m+1$}\label{sub:2l+12m+1}

For this case, the proposed systematic construction method of the
row-monomial DOSTBCs achieving the upper bound of
(\ref{eqn:upperDOSTBCw}) is as follows:
%For convenience, we will uss $\mod(a,b)$ to
%denote the modulo operation, i.e. $\mod(a,b) = a - b\lfloor a/b
%\rfloor$. Furthermore, for two sets $\mathrm{s}_1$ and $\mathrm{s}_2$,
%$\mathrm{s}_1-\mathrm{s}_2$ denotes the set whose elements are in $\mathrm{s}_1$ but
%not in $\mathrm{s}_2$.

Construction IV:

Part I:

{\it Initialization:} Set $p=0$ and
${\cal{S}}=\{s_1,\cdots,s_N\}$.

{\it Step 1:} Neglect $s_{1+ \mod(p,N)}$ and construct a $2 \times
2l$ matrix $\bmX^{(p)}$ in variables ${\cal{S}} - \{s_{1+
\mod(p,N)}\}$ by Construction I.

{\it Step 2:} Set $p=p+1$. If $p <m$, go to {\it Step 1};
otherwise, go to {\it Step 3}.

{\it Step 3:} Let $\bmX_1 =
\left[{\mathrm{diag}}[\bmX^{(0)},\cdots,\bmX^{(m-1)}],\bmzero_{1\times
2lm}\right]^T$ and proceed to Part II.

%Form a $2m \times 2lm$ block diagonal matrix
%$\bar{\bmX}={\mathrm{diag}}[\bmX^{(0)},\cdots,\bmX^{(m-1)}]$. Let
%$\bmX_1 = [\bar{\bmX},\bmzero_{1\times 2lm}]^T$ and proceed to
%Part II.

Part II:

{\it Initialization:} Set $p=0$, ${\cal{S}}^{(K)}={\cal{S}}$, and
$c=1$.
Construct a $K \times \infty$ %\max(2m+l+1,2l+m+1)$
matrix $\bmX_2$ with all zero entries, where $\infty$ means that
the length of $\bmX_2$ is not decided yet.

{\it Step 1:} Set $[\bmX_2]_{2p+1,c}$ equal to
$h_{2p+1}^*s^*_{1+\mod(p,N)}$.

{\it Step 2:} If ${\cal{S}}^{(K)} = \phi$, set $c=c+1$ and go to
{\it Step 4}.

{\it Step 3:} Choose the element with the largest subscript from
${\cal{S}}^{(K)}$ and denote it by $s_{\max}$. Let
$[\bmX_2]_{K,c}$ equal to $h_Ks_{\max}$ and set $c=c+1$. Let
$[\bmX_2]_{2p+1,c}$ and $[\bmX_2]_{K,c}$ equal to
$h_{2p+1}^*s_{\max}^*$ and $-h_Ks_{1+\mod(p,N)}$, respectively.
Set ${\cal{S}}^{(K)} = {\cal{S}}^{(K)} - \{s_{\max},
s_{1+\mod(p,N)}\}$ and $c=c+1$.

{\it Step 4:} Set $p=p+1$. If $p<m$, go to {\it Step 1};
otherwise, set $p=0$ and proceed to {\it Step 5}.

{\it Step 5:} Let $[\bmX_2]_{2p+2,c}$ equal to
$h_{2p+2}s_{1+\mod(p,N)}$.

{\it Step 6:} If ${\cal{S}}^{(K)} = \phi$, set $c=c+1$ and go to
{\it Step 8}.

{\it Step 7:} Choose the element with the largest subscript from
${\cal{S}}^{(K)}$ and denote it by $s_{\max}$. Let
$[\bmX_2]_{K,c}$ equal to $h_K^*s_{\max}^*$ and set $c=c+1$. Let
$[\bmX_2]_{2p+2,c}$ and $[\bmX_2]_{K,c}$ equal to
$-h_{2p+2}s_{\max}$ and $h_K^*s_{1+\mod(p,N)}^*$, respectively.
Set ${\cal{S}}^{(K)} = {\cal{S}}^{(K)} - \{s_{\max},
s_{1+\mod(p,N)}\}$ and $c=c+1$.

{\it Step 8:} Set $p=p+1$. If $p<m$, go to {\it Step 5};
otherwise, discard the all-zero columns at the tail of $\bmX_2$,
build $\bmX(N,K) = [\bmX_1,\bmX_2]$, and end the construction.

For any odd $N\leq 9$ and $K\leq 9$, we have confirmed that the
codes generated by Construction IV achieved the upper bound of
(\ref{eqn:upperDOSTBCw}) indeed. %\footnote{Actually, for $N>10$ and
%$K>4$, we think that the codes may be hard to implement since the
%delay will be very large and the synchronization between relay
%terminals will be very difficult.}
In general, however, it is hard
to prove that Construction IV can generate the row-monomial
DOSTBCs achieving the upper bound of (\ref{eqn:upperDOSTBCw}) for
any odd $N$ and $K$.

For example, when $N=5$ and $K=5$, the matrices $\bmX_1$ and
$\bmX_2$ constructed by Construction IV are given by
\begin{eqnarray}
\bmX_1 &=&\left[%
\begin{array}{cccccccccc}
  h_1s_2 & -h_1s_3 & h_1s_4 & -h_1s_5 &\vline& 0 & 0 & 0 & 0  \\ %&\vline& h_1^*s_1^*&h_1^*s_5^*&0&0\\
  h_2^*s_3^* & h_2^*s_2^* & h_2^*s_5^* & h_2s_4^* &\vline& 0 & 0 & 0 & 0 \\%&\vline&0&0&0&0\\
\hline
  0 & 0 & 0 & 0 &\vline& h_3s_1 & -h_3s_3 & h_3s_4 & -h_3s_5\\%&\vline& 0&0&h_3s_5&0 \\
  0 & 0 & 0 & 0 &\vline& h_4^*s_3^* & h_4^*s_1^* & h_4^*s_5^* & h_4s_4^*\\%&\vline& h_5s_5&-h_5s_1&0&h_4s_5
\hline
  0 & 0 & 0 & 0 & &0 & 0 & 0 & 0 \\
\end{array}%
\right]
\end{eqnarray}
and
\begin{eqnarray}
\bmX_2 &=&\left[%
\begin{array}{cccccccccc}
  h_1^*s_1^*&h_1^*s_5^*&\vline&0&0&\vline&0&0&\vline&0\\
 0&0&\vline&0&0&\vline&h_2s_1&-h_2s_3&\vline&0\\
  0&0&\vline&h_3^*s_2^*&h_3^*s_4^*&\vline&0&0&\vline&0 \\
  0 & 0 &\vline& 0 & 0&\vline&0&0&\vline&h_4s_2\\
  h_5s_5&-h_5s_1&\vline&h_5s_4&-h_5s_2&\vline&h_3^*s_3^*&h_5^*s_1^*&\vline&0 \\
\end{array}%
\right],
\end{eqnarray}
respectively, where the solid lines illustrate the construction
steps. Therefore, $\bmX(5,5) = [\bmX_1,\bmX_2]$ achieves the upper
bound of the data-rate $1/3$.

For this case, the DOSTBCs achieving the upper bound of
(\ref{eqn:rate}) are not found.

%%%%%%%%%%%%%%%%%%%%%%%%%%%%%%%%%%%%%%%%%%%%%%%%%%%%%%%%%%%%%%%%%%%%%%%%%

\section{Numerical Results}\label{sec:num}

In this section, some numerical results are provided to compare
the performance of the DOSTBCs with that of the repetition-based
cooperative strategy. Specifically, we compare the performance of
the codes proposed in Section \ref{sec:code} with that of the
repetition-based strategy. Since the schemes proposed in
\cite{gamal}--\cite{jing2}, \cite{jing3} are not single-symbol ML
decodable, their performance is not compared in this paper.

In order to make the comparison fair, the modulation scheme and
the transmission power per use of every relay terminal need to be
properly chosen for different circumstances. For example, when
$N=4$ and $K=4$, the data-rate of $\bmX(4,4)$ is $1/2$. We choose
Quadrature Phase Shift Keying (QPSK) as the modulation scheme, and
hence, the bandwidth efficiency of $\bmX(4,4)$ is $1$ bps/Hz. On
the other hand, the data-rate of the repetition-based cooperative
strategy is $1/4$. Therefore, $16$-Quadrature Amplitude Modulation
(QAM) is chosen and it makes the bandwidth efficiency of the
repetition-based cooperative strategy $1$ bps/Hz as well.
Similarly, in order to make the bandwidth efficiency equal to $2$
bps/Hz, $16$-QAM and $256$-QAM are chosen for $\bmX(4,4)$ and the
repetition-based cooperative strategy, respectively. Furthermore,
we set the transmission power per use of every relay terminal to
be $E_r$ for $\bmX(4,4)$. Since every relay terminal transmits $4$
times over $8$ time slots, the average transmission power per time
slot is $E_r/2$. For the repetition-based cooperative strategy,
the transmission power per use of every relay terminal is set to
be $2E_r$. Because every relay terminal transmits once over $4$
time slots, the average transmission power per time slot is
$E_r/2$ as well. When $N=4$ and $K=5$, proper modulation schemes
and transmission power per use of every relay terminal can be
found by following the same way.\footnote{When $N=5$ and $K=4$,
the data-rate of $\bmX(5,4)$ is $5/12$. In order to make the
bandwidth efficiency of $\bmX(5,4)$ equal to $1$ bps/Hz or $2$
bps/Hz, the size of the modulation scheme should be $2^{12/5}$ or
$2^{24/5}$, which can not be implemented in practice. Therefore,
we do not evaluate the performance of $\bmX(5,4)$ in this paper.}

When $N=5$ and $K=5$, $8$-PSK and $64$-QAM are chosen for
$\bmX(5,5)$ to make it have the bandwidth efficiency $1$ bps/Hz
and $2$ bps/Hz, respectively. On the other hand, $32$-QAM and
$1024$-QAM are chosen for the repetition-based cooperative
strategy to make it have the bandwidth efficiency $1$ bps/Hz and
$2$ bps/Hz, respectively. For $\bmX(5,5)$, the transmission power
per use of every relay terminal is set to be $E_r$. Because the
fourth relay terminal transmits $5$ times over $15$ time slots,
its average transmission power per time slot is $E_r/3$. Every
other relay terminal transmits $6$ times over $15$ times slots,
and hence, its average transmission power per time slot is
$2E_r/5$. For the repetition-based cooperative strategy, every
relay terminal transmits once over $5$ time slots. Therefore, the
transmission power per use of the fourth relay terminal is set to
be $5E_r/3$, and the transmission power per use of the other relay
terminals is set to be $2E_r$.

%The settings of the modulation scheme and the transmission power
%per use of every relay terminal are summarized
%in Table II for all the considered circumstances. %\ref{tab:num}.

\begin{figure}
\begin{center}
\subpostscript{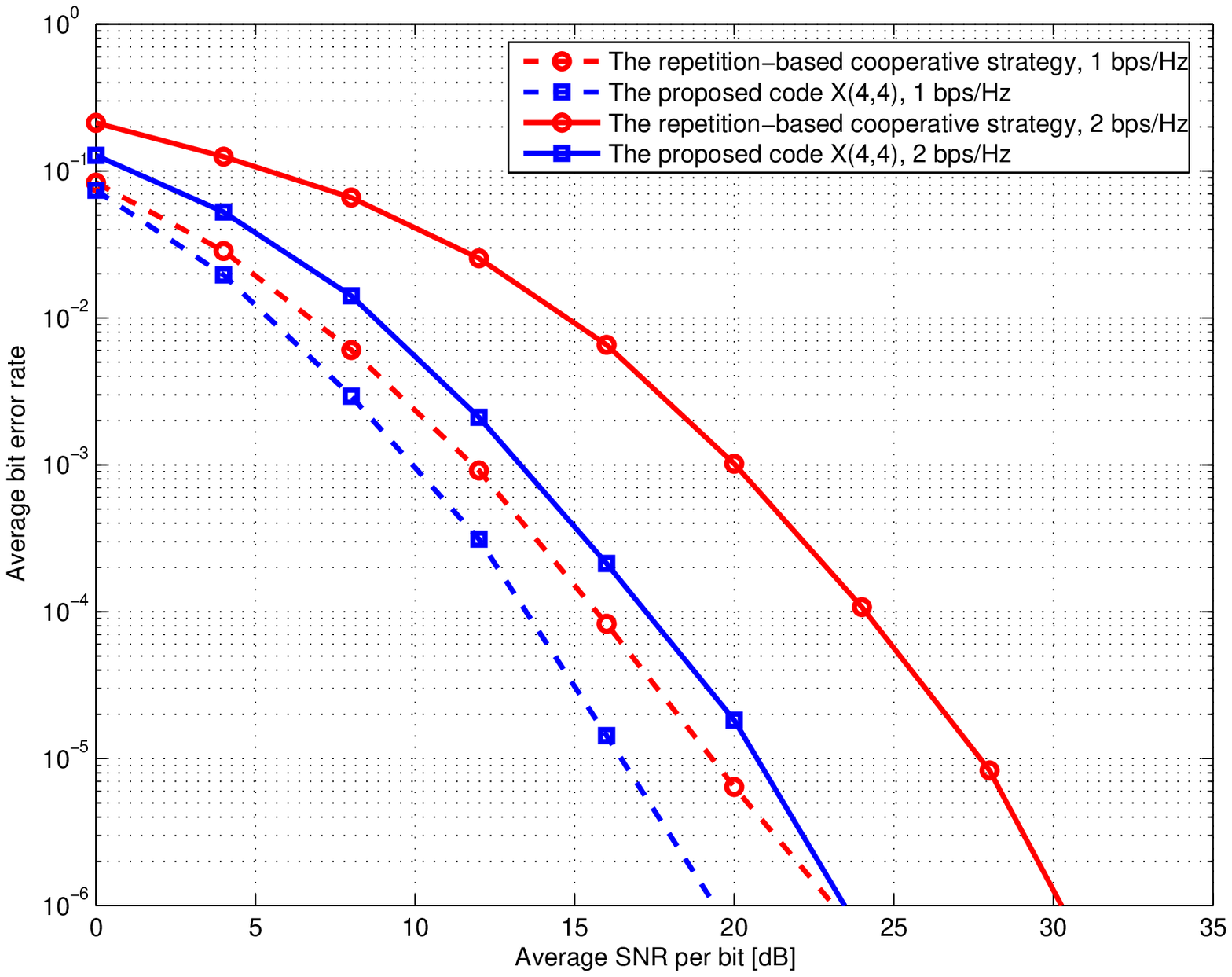}{0.9\textwidth}
\end{center}
\caption{Comparison of the DOSTBCs/row-monomial DOSTBCs with the
repetition-based cooperative strategy, $N=4$, $K=4$.}
\label{fig:n4k4}
\end{figure}

\begin{figure}
\begin{center}
\subpostscript{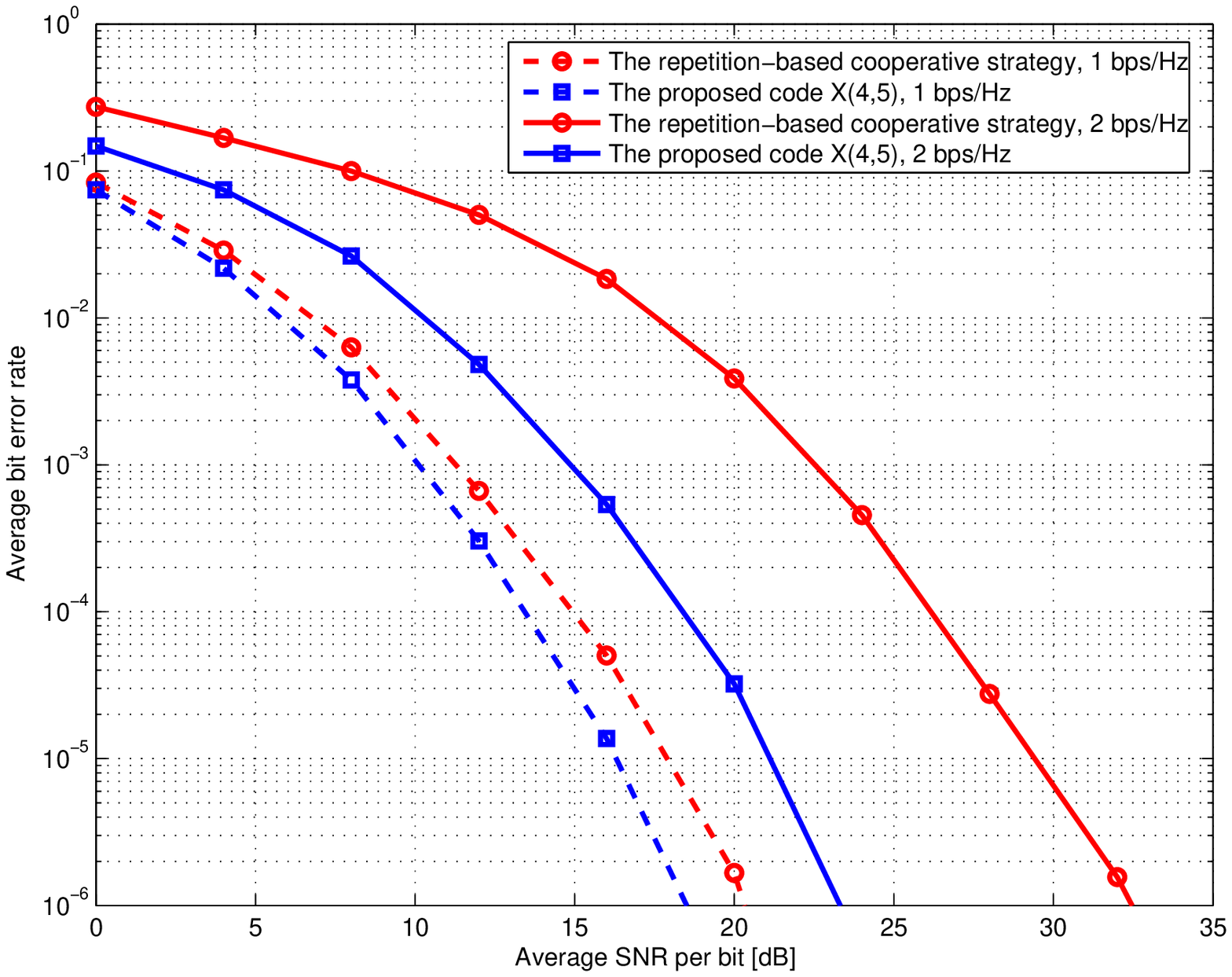}{0.9\textwidth}
\end{center}
\caption{Comparison of the DOSTBCs/row-monomial DOSTBCs with the
repetition-based cooperative strategy, $N=4$, $K=5$.}
\label{fig:n4k5}
\end{figure}

\begin{figure}
\begin{center}
\subpostscript{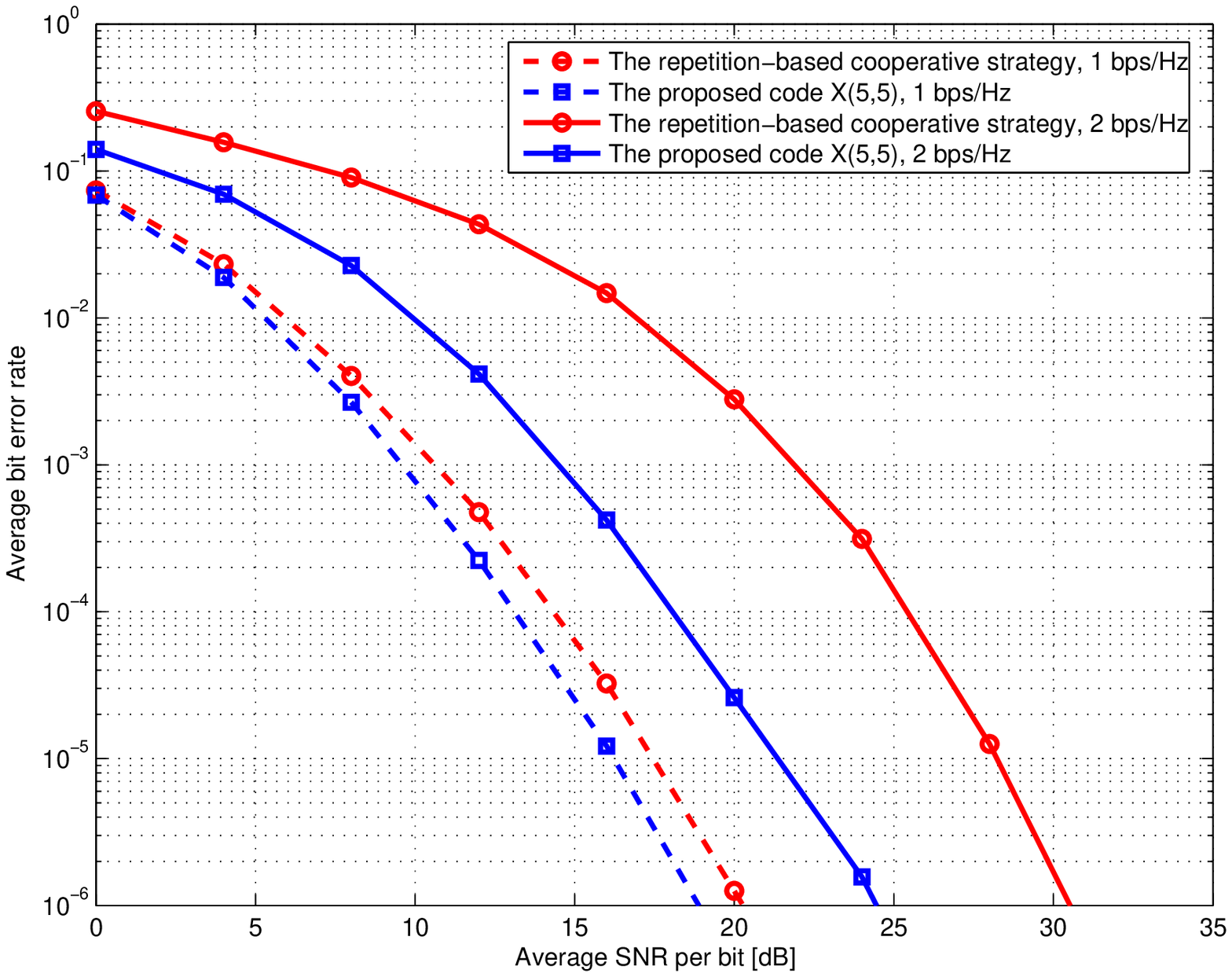}{0.9\textwidth}
\end{center}
\caption{Comparison of the DOSTBCs/row-monomial DOSTBCs with the
repetition-based cooperative strategy, $N=5$, $K=5$.}
\label{fig:n5k5}
\end{figure}

The comparison results are given in Figs.\
\ref{fig:n4k4}--\ref{fig:n5k5}. It can be seen that the
performance of the DOSTBCs is better than that of the
repetition-based cooperative strategy in the whole Signal to Noise
Ratio (SNR) range. The performance gain of the DOSTBCs is more
impressive when the bandwidth efficiency is $2$ bps/Hz. For
example, when $N=4$ and $K=4$, the performance gain of the DOSTBCs
is approximately $7$ dB at $10^{-6}$ Bit Error Rate (BER). This is
because the DOSTBCs have higher data-rate than the
repetition-based cooperative strategy, and hence, they can use the
modulation schemes with smaller constellation size to achieve the
bandwidth efficiency $2$ bps/Hz. When the transmission power is
fixed, smaller constellation size means larger minimum distance
between the constellation points. Since the BER curves shift left
when the minimum distance gets larger, the performance superiority
of the DOSTBCs over the repetition-based cooperative strategy
becomes more obvious when the bandwidth efficiency gets higher.
Furthermore, because the BER curves of the DOSTBCs are parallel
with those of the repetition-based cooperative strategy, they
should have the same diversity order. It is well-known that the
repetition-based cooperative strategy can achieve the full
diversity order $K$. Therefore, the DOSTBCs also achieve the full
diversity order $K$.

\section{Conclusion}\label{sec:conl}

This paper focuses on designing high data-rate distributed
space-time codes with single-symbol ML decodability and full
diversity order. We propose a new type of distributed space-time
codes, DOSTBCs, which are single-symbol decodable and have the
full diversity order $K$. By deriving an upper bound of the
data-rate of the DOSTBC, we show that the DOSTBCs have much better
bandwidth efficiency than the widely used repetition-based
cooperative strategy. However, the DOSTBCs achieving the upper
bound of the data-rate are only found when $N$ and $K$ are both
even. Then, further investigation is given to the row-monomial
DOSTBCs, which result in diagonal noise covariance matrices at the
destination terminal. The upper bound of the data-rate of the
row-monomial DOSTBC is derived and it is equal to or slightly
smaller than that of the DOSTBC. However, systematic construction
methods of the row-monomial DOSTBCs achieving the upper bound of
the data-rate are found for any values of $N$ and $K$. Lastly, the
full diversity order of the DOSTBCs is justified by numerical
results.

%%%%%%%%%%%%%%%%%%%%%%%%%%%%%%%%%%%%%%%%%%%%%%%%%%%%%%%%%%%%%%%%%%%%%

\clearpage
\newpage

\appendices
%\vspace{1cm}
\section*{Appendix A}
\renewcommand\theequation{A.\arabic{equation}}
\setcounter{equation}{0}

\begin{center}
Proof of Lemma \ref{thm:propertyAB}
\end{center}

The first condition is directly from the fact that the entries of
$\bmX$ are 0, $\pm h_k s_n$, $\pm h_k^{*}s_n^{*}$, or multiples of
these indeterminates by $\textbf{j}$.

The proof of the second condition is by contradiction. We assume
that $\bmA_k$ and $\bmB_k$ have non-zero entries at the same
position, for example $[\bmA_k]_{n,t}$ and $[\bmB_k]_{n,t}$ are
both non-zero. Then, $[\bmX]_{k,t}$ will be a linear combination
of $h_k s_n$ and $h_k^{*} s_n^{*}$, which violates the definition
of the DOSTBCs in Definition 1. Thus, $\bmA_k$ and $\bmB_k$ can
not have non-zero entries at the same position.

The proof of the third condition is also by contradiction. In
order to prove $\bmA_k$ is column-monomial, we assume that the
$t$-th, $1\leq t \leq T$, column of it has two non-zero entries:
$[\bmA_k]_{n_1,t}$ and $[\bmA_k]_{n_2,t}$, $n_1\neq n_2$. Then
$[\bmX]_{k,t}$ will be a linear combination of $h_k s_{n_1}$ and
$h_k s_{n_2}$, which violates the definition of the DOSTBCs in
Definition 1. Therefore, any column of $\bmA_k$ can not contain
two non-zero entries. In the same way, it can be easily shown that
any column of $\bmA_k$ can not contain more than two non-zero
entries, and hence, $\bmA_k$ is column-monomial. Similarly, we can
show that $\bmB_k$ is column-monomial.

In order to prove $\bmA_k+\bmB_k$ is column-monomial, we assume
that the $t$-th, $1\leq t \leq T$, column of it contains two
non-zero entries: $[\bmA_k+\bmB_k]_{n_1,t}$ and
$[\bmA_k+\bmB_k]_{n_2,t}$, $n_1\neq n_2$. Because we have already
shown that $\bmA_k$ and $\bmB_k$ can not have non-zero entries at
the same position, there are only two possibilities to make the
assumption hold: 1) $[\bmA_k]_{n_1,t}$ and $[\bmB_k]_{n_2,t}$ are
non-zero; 2) $[\bmA_k]_{n_2,t}$ and $[\bmB_k]_{n_1,t}$ are
non-zero. Under the first possibility, $[\bmX]_{k,t}$ will be a
linear combination of $h_k s_{n_1}$ and $h_k^{*} s_{n_2}^{*}$;
under the second possibility, $[\bmX]_{k,t}$ will be a linear
combination of $h_k s_{n_2}$ and $h_k^{*} s_{n_1}^{*}$. Since both
of them violate the definition of the DOSTBCs in Definition 1, we
can conclude that any column of $\bmA_k+\bmB_k$ can not contain
two non-zero entries. In the same way, it can be easily shown that
any column of $\bmA_k+\bmB_k$ can not contain more than two
non-zero entries, and hence, $\bmA_k+\bmB_k$ is column-monomial.

\appendices
%\vspace{1cm}
\section*{Appendix B}
\renewcommand\theequation{B.\arabic{equation}}
\setcounter{equation}{0}

\begin{center}
Proof of Lemma \ref{thm:originalcond}
\end{center}

The sufficient part is easy to verify. We only prove the necessary
part here, i.e. if (\ref{eqn:orth}) holds, $\bmA_k$ and $\bmB_k$
satisfy (\ref{eqn:cond1})--(\ref{eqn:cond5}).
%Let
%$[\bmX\bmR^{-1}\bmX]_{k_1,k_2}$ denote the $(k_1,k_2)$-th entry of
%$\bmX\bmR^{-1}\bmX$.
When $k_1 \neq k_2$, according to (\ref{eqn:orth}) and
(\ref{eqn:Dn}), $[\bmX\bmR^{-1}\bmX]_{k_1,k_2}$ is given by
\begin{eqnarray}
[\bmX\bmR^{-1}\bmX]_{k_1,k_2} %&=&
%(h_{k_1}\bms\bmA_{k_1}+h^{*}_{k_1}\bms^{*}\bmB_{k_1})\bmR^{-1}(h_{k_2}\bms\bmA_{k_2}
%+h_{k_2}^{*}\bms^{*}\bmB_{k_2})^H\nonumber\\
&=&h_{k_1}h_{k_2}^{*}\bms\bmA_{k_1}\bmR^{-1}\bmA_{k_2}^H\bms^H+
h_{k_1}^*h_{k_2}^{*}\bms^*\bmB_{k_1}\bmR^{-1}\bmA_{k_2}^H\bms^H\nonumber\\
&&+h_{k_1}h_{k_2}\bms\bmA_{k_1}\bmR^{-1}\bmB_{k_2}^H\bms^T+
h_{k_1}^*h_{k_2}\bms^*\bmB_{k_1}\bmR^{-1}\bmB_{k_2}^H\bms^T\nonumber\\ %%\label{eqn:sum1}\\
&=&0.\label{eqn:sum2}
\end{eqnarray}
Note that $h_{k_1}$ and $h_{k_2}$ can be any complex numbers.
Thus, in order to make (\ref{eqn:sum2}) hold for every possible
value of $h_{k_1}$ and $h_{k_2}$, the following equalities must
hold
\begin{eqnarray}
\bms\bmA_{k_1}\bmR^{-1}\bmA_{k_2}^H\bms^H&=&0\\
\bms^*\bmB_{k_1}\bmR^{-1}\bmA_{k_2}^H\bms^H&=&0\\
\bms\bmA_{k_1}\bmR^{-1}\bmB_{k_2}^H\bms^T&=&0\\
\bms^*\bmB_{k_1}\bmR^{-1}\bmB_{k_2}^H\bms^T&=&0.
\end{eqnarray}
By using Lemma 1 of \cite{liang2}, we have (\ref{eqn:cond1}),
(\ref{eqn:cond2}), and
\begin{eqnarray}
\bmA_{k_1} \bmR^{-1} \bmB_{k_2}^H + \bmB_{k_2}^{*}\bmR^{-1}
\bmA_{k_1}^T &=& \bmzero_{N\times N}, \hspace{2cm} 1 \leq k_1\neq k_2 \leq K\label{eqn:cond32}\\
\bmB_{k_1}\bmR^{-1}\bmA_{k_2}^H +
\bmA_{k_2}^*\bmR^{-1}\bmB_{k_1}^T &=& \bmzero_{N\times N},
\hspace{2cm} 1 \leq k_1\neq k_2 \leq K.\label{eqn:cond42}
\end{eqnarray}

When $k_1=k_2=k$, according to (\ref{eqn:orth}) and
(\ref{eqn:Dn}), $[\bmX\bmR^{-1}\bmX]_{k,k}$ is given by
\begin{eqnarray}
[\bmX\bmR^{-1}\bmX]_{k,k} %&=&
%(h_{k}\bms\bmA_{k}+h^{*}_{k}\bms^{*}\bmB_{k})\bmR^{-1}(h_{k}\bms\bmA_{k}
%+h_{k}^{*}\bms^{*}\bmB_{k})^H\nonumber\\
&=&|h_{k}|^2\bms(\bmA_{k}\bmR^{-1}\bmA_{k}^H+
\bmB_{k}^*\bmR^{-1}\bmB_{k}^T)\bms^H\nonumber\\
&&+h_{k}^*h_{k}^{*}\bms^*\bmB_{k}\bmR^{-1}\bmA_{k}^H\bms^H
+h_{k}h_{k}\bms\bmA_{k_1}\bmR^{-1}\bmB_{k}^H\bms^T\nonumber\\
%&=& |h_k|^2\sum_{n=1}^N |s_n|^2 D_{n,k}\\
&=&|h_k|^2\bms \ {\mathrm{diag}}[D_{1,k},\cdots,D_{N,k}]\bms^H.
\end{eqnarray}
For the same reason as in (\ref{eqn:sum2}), the following
equalities must hold
\begin{eqnarray}
\bms(\bmA_{k}\bmR^{-1}\bmA_{k}^H+
\bmB_{k}^*\bmR^{-1}\bmB_{k}^T)\bms^H
&=&\bms \ {\mathrm{diag}}[D_{1,k},\cdots,D_{N,k}]\bms^H\\
\bms^*\bmB_{k}\bmR^{-1}\bmA_{k}^H\bms^H&=&0\\
\bms\bmA_{k_1}\bmR^{-1}\bmB_{k}^H\bms^T&=&0.
\end{eqnarray}
By using Lemma 1 of \cite{liang2} again, we have (\ref{eqn:cond5})
and
\begin{eqnarray}
\bmA_{k} \bmR^{-1} \bmB_{k}^H + \bmB_{k}^{*}\bmR^{-1}
\bmA_{k}^T &=& \bmzero_{N\times N} ,\hspace{2cm} 1 \leq k \leq K\label{eqn:cond33}\\
\bmB_{k}\bmR^{-1}\bmA_{k}^H + \bmA_{k}^*\bmR^{-1}\bmB_{k}^T &=&
\bmzero_{N\times N}, \hspace{2cm} 1 \leq k \leq
K.\label{eqn:cond43}
\end{eqnarray}
Combining (\ref{eqn:cond32}) and (\ref{eqn:cond42}) with
(\ref{eqn:cond33}) and (\ref{eqn:cond43}), respectively, we have
(\ref{eqn:cond3}) and (\ref{eqn:cond4}). This completes the proof
of the necessary part.

\appendices
%\vspace{1cm}
\section*{Appendix C}
\renewcommand\theequation{C.\arabic{equation}}
\setcounter{equation}{0}

\begin{center}
Proof of Theorem \ref{thm:necessary}
\end{center}

If a DOSTBC $\bmX$ exists, (\ref{eqn:orth}) holds by Definition 1%\ref{defi:DOSTBC}
, and hence, (\ref{eqn:cond1})--(\ref{eqn:cond5}) hold by Lemma
\ref{thm:originalcond}. On the other hand, following the same way
of the proof of Lemma \ref{thm:originalcond}, it can be easily
shown that (\ref{eqn:orthE}) holds if and only if
\begin{eqnarray}
\bmA_{k_1}\bmA_{k_2}^H &=& \bmzero_{N\times N}, \hspace{2cm} 1
\leq k_1 \neq k_2 \leq K\label{eqn:Econd1}
\\
\bmB_{k_1}\bmB_{k_2}^H &=& \bmzero_{N\times N}, \hspace{2cm} 1
\leq k_1
\neq k_2 \leq K\label{eqn:Econd2}\\
\bmA_{k_1}  \bmB_{k_2}^H + \bmB_{k_2}^{*}
\bmA_{k_1}^T &=& \bmzero_{N\times N}, \hspace{2cm} 1 \leq k_1,k_2 \leq K\label{eqn:Econd3}\\
\bmB_{k_1}\bmA_{k_2}^H + \bmA_{k_2}^*\bmB_{k_1}^T
&=& \bmzero_{N\times N}, \hspace{2cm} 1 \leq k_1,k_2 \leq K\label{eqn:Econd4}\\
\bmA_k \bmA_k^H + \bmB^{*}_k \bmB^T_ k&=&
{\mathrm{diag}}[E_{1,k},\cdots,E_{N,k}], \hspace{2cm} 1 \leq k
\leq K.\label{eqn:Econd5}
\end{eqnarray}
Therefore, in order to prove Theorem \ref{thm:necessary}, we only
need to show that, if (\ref{eqn:cond1})--(\ref{eqn:cond5}) hold,
(\ref{eqn:Econd1})--(\ref{eqn:Econd5}) hold and $E_{n,k}$ is
strictly positive.

First, we evaluate $[\bmR]_{t_1,t_2}$. According to (\ref{eqn:R}),
when $t_1\neq t_2$, $[\bmR]_{t_1,t_2}$ can be either null or a sum
of several terms containing $|\rho f_k|^2$; when $t_1 = t_2 = t$,
$[\bmR]_{t,t}$ is a sum of a constant 1, which is from the
identity matrix, and several terms containing $|\rho f_k|^2$.
Therefore, we can rewrite $[\bmR]_{t,t}$ as
$[\bmR]_{t,t}=\bar{R}_{t,t}+1$, where $\bar{R}_{t,t}$ accounts for
all the terms containing $|\rho f_k|^2$. $[\bmR^{-1}]_{t_1,t_2}$
is given by $[\bmR^{-1}]_{t_1,t_2} =
C_{t_2,t_1}/\mathrm{det}(\bmR)$, where $C_{t_2,t_1}$ is the matrix
cofactor of $[\bmR]_{t_2,t_1}$. When $t_1 =t_2=t$, by the
definition of matrix cofactor, $C_{t,t}$ contains a constant 1
generated by the product $\prod_{i=1,i\neq t}^T [\bmR]_{i,i} =
\prod_{i=1,i\neq t}^T (\bar{R}_{i,i}+1)$. Furthermore, it is easy
to see that the constant 1 is the only constant term in $C_{t,t}$.
Thus, $C_{t,t}$ can be rewritten as $C_{t,t} = \bar{C}_{t,t}+1$,
where no constant term is in $\bar{C}_{t,t}$. Consequently,
$[\bmR^{-1}]_{t,t}$ can be rewritten as $[\bmR^{-1}]_{t,t} =
\bar{C}_{t,t}/\mathrm{det}(\bmR) + 1/\mathrm{det}(\bmR)$. When
$t_1 \neq t_2$, $C_{t_2,t_1}$ does not contain any constant term,
and hence, $[\bmR^{-1}]_{t_1,t_2}$ does not contain the term
$1/\mathrm{det}(\bmR)$.\footnote{$C_{t_2,t_1}$ may be zero, but it
does not change the conclusion that $[\bmR^{-1}]_{t_1,t_2}$ does
not contain the term $1/\mathrm{det}(\bmR)$.} Therefore, we can
extract the term $1/\mathrm{det}(\bmR)$ from every main diagonal
entry of $\bmR^{-1}$ and rewrite $\bmR^{-1}$ in the following way
\begin{eqnarray}
\bmR^{-1}
&=&\frac{1}{\mathrm{det}(\bmR)}\bar{\bmC}+\frac{1}{\mathrm{det}(\bmR)}\bmI_{T\times
T}.
\end{eqnarray}

Then we show that (\ref{eqn:Econd1}) holds if (\ref{eqn:cond1})
holds. If (\ref{eqn:cond1}) holds, we have
\begin{equation}\label{eqn:arazero}
\bmA_{k_1} \bmR^{-1} \bmA_{k_2}^H  =
\frac{1}{\mathrm{det}(\bmR)}\bmA_{k_1} \bar{\bmC} \bmA_{k_2}^H +
\frac{1}{\mathrm{det}(\bmR)}
\bmA_{k_1}\bmA_{k_2}^H=\bmzero_{N\times N}.
\end{equation}
Note that $\bmR^{-1}$ and $\bar{\bmC}$ are random matrices. In
order to make (\ref{eqn:arazero}) hold for every possible
$\bmR^{-1}$ and $\bar{\bmC}$, both terms in (\ref{eqn:arazero})
must be equal to zero. Therefore, (\ref{eqn:Econd1}) holds.
Similarly, we can show that (\ref{eqn:Econd2})--(\ref{eqn:Econd4})
hold if (\ref{eqn:cond2})--(\ref{eqn:cond4}) hold. Now, we show
that (\ref{eqn:Econd5}) holds if (\ref{eqn:cond5}) holds. If
(\ref{eqn:cond5}) holds, we have
\begin{eqnarray}
\bmA_k \bmR^{-1} \bmA_k^H + \bmB^{*}_k \bmR^{-1} \bmB^T_ k &=&
\frac{1}{\mathrm{det}(\bmR)}\left(\bmA_k \bar{\bmC} \bmA_k^H +
\bmB^{*}_k \bar{\bmC} \bmB^T_ k\right)\nonumber +
\frac{1}{\mathrm{det}(\bmR)}\left(\bmA_k \bmA_k^H +
\bmB^{*}_k\bmB^T_ k
 \right)\\
&=& {\mathrm{diag}}[D_{1,k},\cdots,D_{N,k}].
\end{eqnarray}
For the same reason as in (\ref{eqn:arazero}), the off-diagonal
entries of $\bmA_k \bmA_k^H + \bmB^{*}_k\bmB^T_ k$ must be zero,
and hence, (\ref{eqn:Econd5}) holds.

Lastly, we show that $E_{n,k}$ is strictly positive if
(\ref{eqn:cond5}) holds. From (\ref{eqn:cond5}) and
(\ref{eqn:Econd5}), we have
\begin{eqnarray}
D_{n,k} &=& \sum_{t=1}^T \sum_{i=1}^T [\bmR^{-1}]_{i,t}
([\bmA_k]_{n,i} [\bmA_k]_{n,t}^* + [\bmB_k]_{n,i}^*
[\bmB_k]_{n,t})\\\label{eqn:ARABRAelement2}
 E_{n,k} &=& \sum_{t=1}^T(|[\bmA_k]_{n,t}|^2
+ |[\bmB_k]_{n,t}|^2).\label{eqn:AABBelement2}
\end{eqnarray}
Since $D_{n,k}$ is non-zero, at least one $[\bmA_k]_{n,t}$ or one
$[\bmB_k]_{n,t}$ is non-zero. Furthermore, the modulus of that
non-zero entry is 1 by Lemma \ref{thm:propertyAB}. Therefore,
$E_{n,k}=\sum_{t=1}^T(|[\bmA_k]_{n,t}|^2 + |[\bmB_k]_{n,t}|^2)\geq
1$ is strictly positive, which completes the proof of Theorem
\ref{thm:necessary}.

\appendices
%\vspace{1cm}
\section*{Appendix D}
\renewcommand\theequation{D.\arabic{equation}}
\setcounter{equation}{0}

\begin{center}
Proof of Theorem \ref{thm:upperDOSTBC}
\end{center}

Let $\underline{\bmA} = [\bmA_1,\cdots,\bmA_K]^T$ and
$\underline{\bmB} = [\bmB_1,\cdots,\bmB_K]^T$; then the dimension
of $\underline{\bmA}$ and $\underline{\bmB}$ is $NK\times T$. From
(\ref{eqn:Econd1}), every row of $\bmA_{k_1}$ is orthogonal with
every row of $\bmA_{k_2}$ when $k_1 \neq k_2$.\footnote{A row
vector $\bmx$ is said to be orthogonal with another row vector
$\bmy$ if $\bmx\bmy^H$ is equal to zero.} Furthermore, because
$\bmA_k$ is column-monomial by Lemma \ref{thm:propertyAB}, every
row of $\bmA_k$ is orthogonal with every other row of $\bmA_k$.
Therefore, any two different rows in $\underline{\bmA}$ are
orthogonal with each other, and hence,
${\mathrm{rank}}(\underline{\bmA}) =
\sum_{k=1}^K{\mathrm{rank}}(\bmA_k)$. Similarly, any two different
rows in $\underline{\bmB}$ are orthogonal with each other, and
hence, ${\mathrm{rank}}(\underline{\bmB}) =\sum_{k=1}^K
{\mathrm{rank}}(\bmB_k)$.

On the other hand, by (\ref{eqn:Econd5}), we have
\begin{equation}
{\mathrm{rank}}(\bmA_k) +{\mathrm{rank}}(\bmB_k) \geq
{\mathrm{rank}}({\mathrm{diag}}[E_{1,k},\cdots,E_{N,k}]) =  N,
\end{equation}
where the inequality is from the rank inequality 3) in
\cite{wang}, and hence,
\begin{eqnarray}
\sum_{k=1}^K {\mathrm{rank}}(\bmA_k) + \sum_{k=1}^K
{\mathrm{rank}}(\bmB_k) \geq NK.
\end{eqnarray}
Because ${\mathrm{rank}}(\underline{\bmA})$ and
${\mathrm{rank}}(\underline{\bmB})$ are integers, we have
\begin{equation}\label{eqn:rankA}
{\mathrm{rank}}(\underline{\bmA}) = \sum_{k=1}^K
{\mathrm{rank}}(\bmA_k) \geq \left\lceil \frac{NK}{2}\right\rceil
\end{equation}
or
\begin{equation}\label{eqn:rankB}
{\mathrm{rank}}(\underline{\bmB}) = \sum_{k=1}^K
{\mathrm{rank}}(\bmB_k) \geq \left\lceil \frac{NK}{2}\right\rceil.
\end{equation}
If (\ref{eqn:rankA}) is true, $T\geq
{\mathrm{rank}}(\underline{\bmA}) \geq \left\lceil
(NK)/2\right\rceil$ and (\ref{eqn:rate}) holds. If
(\ref{eqn:rankB}) is true, the same conclusion can be made.

\appendices
%\vspace{1cm}
\section*{Appendix E}
\renewcommand\theequation{E.\arabic{equation}}
\setcounter{equation}{0}

\begin{center}
Proof of Theorem \ref{thm:whitenoise}
\end{center}

The sufficient part is easy to verify. We only prove the necessary
part here, i.e. if $\bmR$ is a diagonal matrix, $\bmA_k$ and
$\bmB_k$ are row-monomial. This is done by contradiction. If
$\bmR$ is a diagonal matrix, the off-diagonal entries
$[\bmR]_{t_1,t_2}$, $1\leq t_1\neq t_2\leq T$, are equal to 0.
According to (\ref{eqn:R}), we have
\begin{equation}
[\bmR]_{t_1,t_2} = \sum_{k=1}^K \left[|\rho
f_k|^2\left(\sum_{n=1}^N [\bmA_k]_{n,t_1}^*[\bmA_k]_{n,t_2} +
\sum_{n=1}^N [\bmB_k]_{n,t_1}^*[\bmB_k]_{n,t_2}\right) \right]=0.
\end{equation}
In order to make the equality hold for every possible $f_k$, the
following equality must hold
\begin{eqnarray}\label{eqn:zerosum}
\sum_{n=1}^N [\bmA_k]_{n,t_1}^*[\bmA_k]_{n,t_2} + \sum_{n=1}^N
[\bmB_k]_{n,t_1}^*[\bmB_k]_{n,t_2} &=&0, \hspace{1.5cm} 1\leq k
\leq K.
\end{eqnarray}
Let us assume that the $n^{'}$-th, $1\leq n^{'} \leq N$, row of
$\bmA_k$ contains two non-zero entries: $[\bmA_k]_{n^{'},t_1}$ and
$[\bmA_k]_{n^{'},t_2}$, $1\leq t_1\neq t_2\leq T$. Because
$\bmA_k$ is column-monomial according to Lemma
\ref{thm:propertyAB}, $[\bmA_k]_{n,t_1}=[\bmA_k]_{n,t_2}=0$,
$1\leq n\neq n^{'}\leq N$, and hance,
\begin{equation}\label{eqn:sumaa}
\sum_{n=1}^N [\bmA_k]_{n,t_1}^*[\bmA_k]_{n,t_2} =
[\bmA_k]_{n^{'},t_1}^* [\bmA_k]_{n^{'},t_2}\neq 0.
\end{equation}
On the other hand, because $\bmA_k$ and $\bmB_k$ can not have
non-zero entries at the same place according to Lemma
\ref{thm:propertyAB}, we have
$[\bmB_k]_{n^{'},t_1}=[\bmB_k]_{n^{'},t_2}=0$. Furthermore,
because $\bmA_k+\bmB_k$ is column-monomial, $[\bmB_k]_{n,t_1} =
[\bmB_k]_{n,t_2}=0$, $1\leq n \neq n^{'} \leq N$. Therefore,
$[\bmB_k]_{n,t_1} = [\bmB_k]_{n,t_2}=0$, $1\leq n \leq N$, and
consequently, $\sum_{n=1}^N [\bmB_k]_{n,t_1}^*[\bmB_k]_{n,t_2}
=0$. It follows from (\ref{eqn:zerosum}) and $\sum_{n=1}^N
[\bmB_k]_{n,t_1}^*[\bmB_k]_{n,t_2} =0$ that
\begin{eqnarray}\label{eqn:sumaazero}
\sum_{n=1}^N [\bmA_k]_{n,t_1}^*[\bmA_k]_{n,t_2} = 0.
\end{eqnarray}
Because (\ref{eqn:sumaa}) and (\ref{eqn:sumaazero}) contradict
with each other, we can conclude that any row of $\bmA_k$ can not
contain two non-zero entries. Furthermore, in the same way, it can
be easily shown that any row of $\bmA_k$ can not contain more than
two non-zero entries, and hence, $\bmA_k$ is row-monomial.
Similarly, we can show that $\bmB_k$ is row-monomial, which
completes the proof of the necessary part.

%The first condition is directly from Lemma
%\ref{thm:rowmonomial}: when either the $n$-th row of $\bmA_k$ or
%the $n$-th row of $\bmB_k$ contains one non-zero entry, the value
%of $E_{nk}$ is 1; when both the $n$-th row of $\bmA_k$ and the
%$n$-th row of $\bmB_k$ contain one non-zero entry, the value of
%$E_{nk}$ is 2.

\appendices
%\vspace{1cm}
\section*{Appendix F}
\renewcommand\theequation{F.\arabic{equation}}
\setcounter{equation}{0}

\begin{center}
Proof of Lemma \ref{thm:rowmonomial}
\end{center}

The proof is by contradiction. We assume that the $t^{'}$-th
column of $\bmA_{k_1}$ and $\bmA_{k_2}$, $1\leq k_1\neq k_2 \leq
K$, contains a non-zero entry $[\bmA_{k_1}]_{n_1,t^{'}}$ and a
non-zero entry $[\bmA_{k_2}]_{n_2,t^{'}}$, respectively. By
Definition 2, $\bmA_{k_1}$ and $\bmA_{k_2}$ are both row-monomial,
we have $[\bmA_{k_1}]_{n_1,t} = [\bmA_{k_2}]_{n_2,t}=0$, $1\leq
t\neq t^{'}\leq T$, and hence,
\begin{equation}\label{eqn:AAdisjoint1}
\sum_{t=1}^T [\bmA_{k_1}]_{n_1,t} [\bmA_{k_2}]_{n_2,t}^* =
[\bmA_{k_1}]_{n_1,t^{'}} [\bmA_{k_2}]_{n_2,t^{'}}^*\neq 0.
\end{equation}
On the other hand, from (\ref{eqn:Econd1}),
$[\bmA_{k_1}\bmA_{k_2}^H]_{n_1,n_2}$ is given by
\begin{eqnarray}\label{eqn:AAdisjoint2}
[\bmA_{k_1}\bmA_{k_2}^H]_{n_1,n_2}=\sum_{t=1}^T
[\bmA_{k_1}]_{n_1,t} [\bmA_{k_2}]_{n_2,t}^* =0.
\end{eqnarray}
Because (\ref{eqn:AAdisjoint1}) and (\ref{eqn:AAdisjoint2})
contradict with each other, we conclude that $\bmA_{k_1}$ and
$\bmA_{k_2}$, $1\leq k_1\neq k_2 \leq K$, can not contain non-zero
entries on the same column simultaneously. Therefore, $\bmA_{k_1}$
and $\bmA_{k_2}$ are column-disjoint when $k_1 \neq k_2$.
Similarly, we can show that $\bmB_{k_1}$ and $\bmB_{k_2}$ are
column-disjoint when $k_1 \neq k_2$.

\appendices
%\vspace{1cm}
\section*{Appendix G}
\renewcommand\theequation{G.\arabic{equation}}
\setcounter{equation}{0}

\begin{center}
Proof of Lemma \ref{thm:tpyeii}
\end{center}

If no Type-II column exists in $\bmX$, it is trivial that the
number of the Type-II columns in $\bmX$ is even. If there is one
Type-II column in $\bmX$, without loss of generality, we assume
that the $t_1$-th column in $\bmX$ is a Type-II column and it
contains $h_{k_1} s_{n_1}$ and $h_{k_2}^* s_{n_2}^*$ on the
$k_1$-th and $k_2$-th row, respectively. Consequently, the inner
product of the $k_1$-th row and the $k_2$-th row will contain the
term $h_{k_1} s_{n_1}h_{k_2} s_{n_2}$.\footnote{The inner product
of two row vectors $\bmx$ and $\bmy$ is defined as $\bmx\bmy^H$.}
Because $\bmX$ is a row-monomial DOSTBC, the inner product of any
two different rows is null by (\ref{eqn:orthE}). Hence, the inner
product of the $k_1$-th row and the $k_2$-th row should contain
the term $-h_{k_1} s_{n_1}h_{k_2} s_{n_2}$ as well to cancel the
term $h_{k_1} s_{n_1}h_{k_2} s_{n_2}$. Thus, there must be another
Type-II column, for example the $t_2$-th column, $t_1\neq t_2$,
which contains $-h_{k_1} s_{n_2}$ and $h_{k_2}^* s_{n_1}^*$ on the
$k_1$-th and $k_2$-th row, respectively. Therefore, the Type-II
columns in $\bmX$ always appear in pairs, and hence, the total
number of the Type-II columns in $\bmX$ is even.

For convenience, we will refer to any entry in $\bmX$ that
contains $s_n$ or $s_n^*$ as the $s_n$-entry. If no $s_n$-entry
exists in the Type-II columns of $\bmX$, it is trivial that the
total number of $s_n$-entries in the Type-II columns of $\bmX$ is
even. If there is one $s_n$-entry in a Type-II column of $\bmX$,
we assume it contains $s_n$ without loss of generality. From the
proof of the first property in Lemma \ref{thm:tpyeii}, we can see
that there must be an $s_n$-entry in another Type-II column and it
contains $s_n^*$. Therefore, in the Type-II columns of $\bmX$, the
$s_n$-entries always appear in pairs, and hence, the total number
of the $s_n$-entries in the Type-II columns of $\bmX$ is even.

\appendices
%\vspace{1cm}
\section*{Appendix H}
\renewcommand\theequation{H.\arabic{equation}}
\setcounter{equation}{0}

\begin{center}
Proof of Theorem \ref{thm:upperDOSTBCw}
\end{center}

For convenience, we will refer to any entry in $\bmX$ that
contains $s_n$ or $s_n^*$ as the $s_n$-entry. Let $U$ denote the
total number of non-zero entries in $\bmX$; $V_n$ the total number
of $s_n$-entries in $\bmX$; $W_k$ the total number of non-zero
entries in the $k$-th row of $\bmX$. Obviously, $U = \sum_{n=1}^N
V_n = \sum_{k=1}^K W_k$. According to (\ref{eqn:Econd5}) and
(\ref{eqn:AABBelement2}), at least one $[\bmA_k]_{n,t}$ or one
$[\bmB_k]_{n,t}$ is non-zero, $1\leq n \leq N$ and $1\leq k\leq
K$. Thus, every row of $\bmX$ has at least one $s_n$-entry, $1\leq
n \leq N$. On the other hand, by the row-monomial condition of
$\bmA_k$ and $\bmB_k$, every row of $\bmX$ has at most two
$s_n$-entries, where one contains $s_n$ and the other contains
$s_n^*$. Therefore, every row of $\bmX$ contains at least $N$ and
at most $2N$ non-zero entries, i.e. $N\leq W_k \leq 2N$, $1\leq k
\leq K$. For the same reason, we have $K \leq V_n \leq 2K$, $1\leq
n \leq N$. Consequently, we have $NK \leq U \leq 2NK$.

{\it Case I: $N=2l$ and $K=2m$.} When $N=2l$ and $K=2m$, $U\geq NK
= 4lm$. Because a pair of Type-II columns contains 4 non-zero
entries, at least $\left\lceil 4lm/4 \right\rceil$ pairs of
Type-II columns are needed to transmit all the non-zero entries.
Since $T$ is the total number of columns in $\bmX$, we have the
following inequality
\begin{equation}
T \geq 2\left\lceil \frac{4lm}{4} \right\rceil = 2lm,
\end{equation}
and hence,
\begin{eqnarray}
{\mathrm{Rate}}_r &\leq& \frac{1}{m}.
\end{eqnarray}
%In Section \ref{sec:code}, we will design the row-monomial DOSTBC that
%reaches the highest rate.

{\it Case II: $N=2l+1$ and $K=2m$.} When $N=2l+1$ and $K=2m$,
without loss of generality, we assume $W_1,\cdots,W_w$ are even
and $W_{1+w},\cdots,W_{2m}$ are odd, where $1\leq w \leq 2m$. We
first have $U\geq NK = 4lm+2m$. Furthermore, because $W_k$ is even
for $1\leq k\leq w$, $W_k\geq N+1= 2l+2$. Consequently, $U\geq 4lm
+2m +w$. On the other hand, because the Type-II columns always
appear in pairs, the $k$-th row of $\bmX$, $w+1 \leq k \leq 2m$,
must contain at least one Type-I column; otherwise, $W_k$ will be
even, which violates our assumption. Therefore, there are at least
$2m-w$ Type-I columns in $\bmX$ and they contain $2m-w$ non-zero
entries. Because a pair of Type-II columns contains 4 non-zero
entries, the rest non-zero entries need at least $\left\lceil (4lm
+2m +w-(2m-w))/4 \right\rceil$ pairs of Type-II columns to
transmit. Therefore, we have the following inequality
\begin{eqnarray}
T&\geq& 2m-w +2\left\lceil \frac{4lm +2m +w-(2m-w)}{4} \right\rceil\\
&\geq& 2m-w + \frac{4lm +2m +w-(2m-w)}{2}\\
%&\geq& 2m-w + \frac{4lm+2m+w-(2m-w)}{2}\\
%&=&2m-w+2lm+w\\
&=&2lm+2m,
\end{eqnarray}
and hence,
\begin{eqnarray}
{\mathrm{Rate}}_r &\leq& \frac{2l+1}{2lm+2m}.
\end{eqnarray}

{\it Case III: $N=2l$ and $K=2m+1$.} When $N=2l$ and $K=2m+1$,
without loss of generality, we assume $V_1,\cdots, V_v$ are even
and $V_{v+1},\cdots,V_{2l}$ are odd, where $1\leq v \leq 2l$. We
first have $U\geq NK = 4lm+2l$. Furthermore, because $V_n$ is even
for $1\leq n\leq v$, $V_n \geq K+1 = 2m+2$. Consequently, $U\geq
4lm+2l+v$. On the other hand, because the total number of
$s_n$-entries in the Type-II columns of $\bmX$ is even, at least
one $s_n$-entry, $v+1\leq n \leq 2l$, is in a Type-I column;
otherwise, $V_n$ will be even, which violates our assumption.
Thus, there are at least $2l-v$ Type-I columns in $\bmX$ and they
contain $2l-v$ non-zero entries. Because a pair of Type-II columns
contains 4 non-zero entries, the rest non-zero entries need at
least $\left\lceil (4lm+2l+v-(2l-v))/4 \right\rceil$ pairs of
Type-II columns to transmit. Therefore, we have the following
inequality
\begin{eqnarray}
T&\geq& 2l-v +2\left\lceil \frac{4lm+2l+v-(2l-v)}{4} \right\rceil\\
&\geq& 2l-v + \frac{4lm+2l+v-(2l-v)}{2}\\
%&\geq& 2l-v + \frac{4lm+2l+v-(2l-v)}{2}\\
%&=&2l-v+2lm+v\\
&=&2lm+2l,
\end{eqnarray}
and hence,
\begin{eqnarray}
{\mathrm{Rate}}_r &\leq& \frac{1}{m+1}.
\end{eqnarray}

{\it Case IV: $N=2l+1$ and $K=2m+1$.} When $N=2l+1$ and $K=2m+1$,
we can assume that $W_1,\cdots,W_w$ are even and
$W_{1+w},\cdots,W_{2m+1}$ are odd, where $1\leq w \leq 2m+1$. By
following the proof of Case II, we have
\begin{eqnarray}
T&\geq&2m+1-w +2\left\lceil \frac{4lm+2l+2m+1+w-(2m+1-w)}{4}\right\rceil \\
%&\geq&2m+1-w+\frac{U-(2m+1-w)}{2}\\
&\geq&2m+1-w+\frac{4lm+2l+2m+1+w-(2m+1-w)}{2}\\
%&=&2m+1-w+2lm+l+w\\
&=&2lm+2m+l+1.\label{eqn:4eq1}
\end{eqnarray}
On the other hand, we can assume $V_1,\cdots, V_v$ are even and
$V_{v+1},\cdots,V_{2l+1}$ are odd, where $1\leq v \leq 2l+1$. By
following the proof of Case III, we have
\begin{eqnarray}
T&\geq& 2l+1-v +2\left\lceil \frac{4lm+2l+2m+1+v-(2l+1-v)}{4} \right\rceil\\
%&\geq& 2l+1-v + \frac{U-(2l+1-v)}{2}\\
&\geq& 2l+1-v + \frac{4lm+2l+2m+1+v-(2l+1-v)}{2}\\
%&=&2l+1-v+2lm+m+v\\
&=&2lm+2l+m+1.\label{eqn:4eq2}
\end{eqnarray}
From (\ref{eqn:4eq1}) and (\ref{eqn:4eq2}), it is immediate that
\begin{eqnarray}
T&\geq&\max(2lm+2m+l+1,2lm+2l+m+1),
\end{eqnarray}
and
\begin{eqnarray}
{\mathrm{Rate}}_r &\leq& \min\left(\frac{2l+1}{2lm+2m+l+1},
\frac{2l+1}{2lm+2l+m+1}\right).
\end{eqnarray}

%%%%%%%%%%%%%%%%%%%%%%%%%%%%%%%%%%%%%%%%%%%%%%%%5

%\input{symbol_decode}

\clearpage
\newpage

\listoffigures

\listoftables

\clearpage
\newpage

%\begin{figure}
%\begin{center}
%\subpostscript{N5K4.eps}{0.9\textwidth}
%\end{center}
%\caption{Comparison of the DOSTBCs/row-monomial DOSTBCs with the
%repetition-based cooperative strategy, $N=5$, $K=4$.}
%\label{fig:n5k4}
%\end{figure}

\newpage
\clearpage

\begin{table}[h]
\begin{center}\label{tab:bounds}
\caption{Upper Bounds of the Data-Rates of the DOSTBC and
Row-Monomial DOSTBC}
\begin{tabular}{|c|c|c|c|}
  \hline
   & DOSTBCs & row-monomial DOSTBCs & difference \\
  \hline
  $N=2l$, $K=2m$ & $\frac{1}{m}$  & $\frac{1}{m}$ & 0 \\
  \hline
  $N=2l+1$, $K=2m$ & $\frac{1}{m}$ & $\frac{2l+1}{2lm+2m}$ & $\frac{1}{2lm+2m}$ \\
   \hline
  $N=2l$, $K=2m+1$  & $\frac{2}{2m+1}$ & $\frac{1}{1+m}$ & $\frac{1}{(2m+1)(m+1)}$\\
   \hline
  \multirow{2}{*}{$N=2l+1$, $K=2m+1$} & \multirow{2}{*}{$\frac{2l+1}{2lm+l+m+1}$} &
  $\min\left(\frac{2l+1}{2lm+2m+l+1}\right.,$
  & $\max\left(\frac{m(2l+1)}{(2lm+l+m+1)(2lm+2m+l+1)}\right.,$ \\
  & &\hspace{0.5cm}$\left.\frac{2l+1}{2lm+2l+m+1}\right)$
  & \hspace{0.6cm}$\left.\frac{l(2l+1)}{(2lm+l+m+1)(2lm+2l+m+1)}\right)$ \\
  \hline
\end{tabular}
\end{center}
\end{table}

\end{document}